\documentclass[aps,pre,superscriptaddress,twocolumn,preprintnumbers]{revtex4-2}

\usepackage{amsmath}
\usepackage{amssymb}
\usepackage{xspace}
\usepackage{xcolor}
\usepackage{graphicx}
\usepackage{dsfont}
\usepackage{setspace}

\newcommand{\simulation}{\widetilde{\Phi}^\tau_\alpha}
\newcommand{\realization}{\ldots, x_{-1}, x_0, x_1, \ldots}
\newcommand{\dynproc}{\{ \realization \}}
\newcommand{\timeseries}{\{x_0, x_1, \ldots, x_T\}}
\newcommand{\stocproc}{\{ \dots, X_{-1}, X_0, X_1, \dots \}}

\newcommand{\iddmodel}{\mathcal{T}^\tau}

\newcommand{\edmd}{\mathrm{\mathbf{U}}_X^\tau}

\newcommand{\dict}{\mathbf{\Psi}}
\newcommand{\partition}{\mathbb{P}}
\newcommand{\Xp}{X_\partition}

\newcommand{\past}{\overleftarrow{x}}
\newcommand{\Past}{\overleftarrow{X}}

\newcommand{\markovOP}{\overleftarrow{M}_0}

\DeclareMathOperator*{\argmin}{argmin}

\begin{document}



\title{Nonequilibrium Statistical Mechanics and\\
Optimal Prediction of\\
Partially-Observed Complex Systems}



\author{Adam Rupe}
\email[]{adamrupe@lanl.gov}
\affiliation{Center For Nonlinear Studies, Theory Division, Los Alamos National Laboratory}
\affiliation{Computational Earth Science, Earth and Environmental Sciences Division, Los Alamos National Laboratory}


\author{Velimir V. Vesselinov}
\email[]{vvv@lanl.gov}
\affiliation{Computational Earth Science, Earth and Environmental Sciences Division, Los Alamos National Laboratory}

\author{James P. Crutchfield}
\email[]{chaos@ucdavis.edu}
\affiliation{Complexity Sciences Center, Department of Physics and Astronomy, University of California Davis}

\date{\today}

\begin{abstract}
Only a subset of degrees of freedom are typically accessible or measurable in
real-world systems. As a consequence, the proper setting for empirical modeling
is that of partially-observed systems. Notably, data-driven models consistently
outperform physics-based models for systems with few observable degrees of
freedom; e.g., hydrological systems. Here, we provide an operator-theoretic
explanation for this empirical success. To predict a partially-observed
system's future behavior with physics-based models, the missing degrees of
freedom must be explicitly accounted for using data assimilation and model
parametrization. Data-driven models, in contrast, employ delay-coordinate
embeddings and their evolution under the Koopman operator to implicitly model
the effects of the missing degrees of freedom. We describe in detail the
statistical physics of partial observations underlying data-driven models using
novel Maximum Entropy and Maximum Caliber measures. The resulting
nonequilibrium Wiener projections applied to the Mori-Zwanzig formalism reveal
how data-driven models may converge to the true dynamics of the observable degrees
of freedom. Additionally, this framework shows how data-driven models infer the effects of unobserved degrees of freedom implicitly, in much the same way that physics models infer the effects explicitly. This provides a unified implicit-explicit modeling framework for predicting partially-observed systems, with hybrid physics-informed machine learning methods combining implicit and explicit aspects.
\end{abstract}


\maketitle
\begin{spacing}{0.9}
\end{spacing}

\section{Introduction}

Most Earth Science investigations access only a subset of a high-dimensional
dynamical system's degrees of freedom due to limited instrumentation.
Predicting the future behavior of partially-observed systems is a central
challenge for many areas of Earth Science, and one that dates back to the
earliest uses of scientific computing \cite{edwa10a,dyso12a}.

Traditional prediction employing \emph{physics-based} (or \emph{process-based})
\emph{models} relies on \emph{explicit representations}: systems are modeled via
closed-form equations of motion that determine how a system evolves forward in
time through interactions among all its degrees of freedom. Predictions are extracted from numerical approximations of solutions of
the equations of motion. This requires knowing the full state of the system at
each time, but limited instrument measurements of the true system provide only
a partial view of the underlying state. \emph{Data assimilation} is then used
to generate a \emph{data-image} through model inversion. The result is a
coarse-grained approximation of the full system state that is most consistent
with the instrument observations and assumptions of the underlying physics.

In contrast, \emph{data-driven prediction} (typically) does not rely on explicit
closed-form models and thus does not require interpolated data-images. For the
prediction task that evolves only instrument measurements forward in time,
data-driven models learn \emph{implicit representations} for this evolution
directly from the observations themselves.

The explicit nature of physics models hinges on our understanding of the
underlying physics governing the system being encapsulated in closed-form
differential equations-of-motion. This is what explicit representations attempt
to approximate. The equivalent governing physics---the ``ground truth''---for
the evolution of the measurement observables is given by linear,
infinite-dimensional Koopman operators. The implicit representations of data-driven models thus attempt to
learn projections of the Koopman operators' action~\cite{berr20a}.

In fact, the governing equations of motion for the measurement observables are
given by the Mori-Zwanzig equation \cite{Wild98a,chor02a}, derived from
expanding the action of the Koopman operator in terms of projection operators
onto the observable degrees of freedom \cite{lin21a,lin21b}. A key insight from
the Mori-Zwanzig formalism is that predictive models of partially-observed
systems require a history dependence---past observations of the observable
degrees of freedom generally contain information relevant for future
predictions.



Recently, the connection between the history dependence of predictive models
and the intrinsic geometry of delay-coordinate embeddings \cite{Pack80,Take81}
has been explored \cite{brun17a,arba17a,gian19a,kamb20a}. Past values of
partial observations, in the form of delay embeddings, implicitly stand in for
the missing degrees of freedom. This parallels how, for physics-based
models, data images act explicitly to fill in the gaps of the missing
degrees of freedom when predicting partially-observed systems.

Most data-driven modeling and prediction relies on Hilbert space methods that
learn a \emph{target function} living in a Hilbert space of functions \cite{alex20a}. Optimal Hilbert space models take the form of a \emph{conditional expectation} of future observations given past observations. This optimum is equivalent to a nonlinear projection of the action of the Koopman operator (that gives the future value of the measurement observables) onto the Hilbert subspace of functions of only the observable degrees of freedom. History-dependent target function models can be expressed as functions of delay-coordinate embeddings using Wiener projections \cite{lin21a}. The optimal model is then the nonlinear Wiener projection of the action of the Koopman operator onto functions of observed delay embeddings.

There is evidence that Wiener projection models may converge to the true
dynamics of the measurement observables if sufficient past observations are
taken into account \cite{gila21a}. Here, we provide a new perspective on the
behavior of history-dependent data-driven models and their relation to the true
underlying physics of partial observations. We do so using insights from the
logical inference approach to statistical mechanics given by Jaynes' Maximum
Entropy principle \cite{Jayn57a}. This further builds on the connections
between nonequilibrium statistical mechanics and optimal prediction of
partially-observed systems \cite{chor00a}.

Optimal Hilbert space models are typically formulated in terms of an invariant
``equilibrium'' measure. However, we show there is a natural family of
time-dependent ``nonequilibrium'' measures induced by partial observations
using Maximum Entropy and its time-varying generalization Maximum Caliber
\cite{Jayn85a,Gran08a}. Constructively, these measures support more general
nonasymptotic behaviors---behaviors that cannot be modeled with an invariant
measure. Importantly, though, they provide unique insights into the convergence
of optimal models to the true governing physics of partial observations. They
do this by directly constructing \emph{predictive
distributions}---probabilities over future observations given past
observations. In particular, we express the possible convergence of
history-dependent models as a thermodynamic limit in which the variance of
predictive distributions vanishes as the length of past observations increases.
This again shows how the action of Koopman operators on delay embeddings implicitly account for the effects of
unobserved degrees of freedom.

Formulating optimal data-driven models as expectations of predictive
distributions suggests a more general stochastic framework for modeling partially-observed
systems. Rather than returning the expectation of predictive distributions,
optimal stochastic models simply return the predictive distributions themselves
\cite{Shal98a}, which then may be sampled for ensemble forecasts. Our direct
construction of predictive distributions using Maximum Caliber measures leads
naturally to such optimal stochastic models for partially-observed systems. A
sequel gives this stochastic formulation of optimal prediction of
partially-observed systems.

\subsection{Implicit versus explicit representations}

The physical insights that emerge shed light on why data-driven models can
outperform traditional physics models for predicting systems with relatively
few observed degrees of freedom. Indeed, this has become increasingly common
for hydrological systems \cite{krat18a,krat19a,read19a,jia21a}. In these cases,
the implicit approach that uses delay-coordinate embeddings is more effective
than the explicit approach that uses data assimilation. For example, while many
details, e.g., subsurface morphology, are crucial for geophysical prediction,
given limited available subsurface measurements, reconstructing informative
data-images for them is exceedingly difficult. This leads to less effective
physics-based methods that rely on the latter.

Perhaps unsurprisingly in this light, due to their empirical successes in
scientific applications, data-driven predictive models are increasingly
employed. That said, they are widely considered to be an entirely new
paradigm---a paradigm with little to no relation with governing physics and
physics-based models. We aim to show that they are in fact quite similar. 

Our framework, together with numerical examples,
shows that data-driven models do implicitly what physics-based models do
explicitly to account for unobserved degrees of freedom. 
We also clarify how the action Koopman and Perron-Frobenius operators on delay-coordinate embeddings may converge to the true system dynamics on the full  system state. 
Generating partitions on maps of the unit interval are discussed as a rigorous example displaying this behavior.
Said another way, the physics underlying history-dependent data-driven models is the same as the physics underlying traditional physics-based models. 

The resulting unified modeling framework shows that the distinction is not so
much ``data-driven versus physics-based'', but rather the emphasis should be on
where approaches land in the ``implicit versus explicit'' representation
spectrum. The class of \emph{physics-informed machine learning} models
\cite{will20a,karn21a,kash21a}, now rapidly gaining popularity, are thus seen
to lie between fully-explicit physics-based models and fully-implicit
data-driven models. Such hybrid models explicitly enforce certain physical
properties as \emph{inductive biases}
\cite{batt18a,bron21a}, with any remaining properties
learned implicitly from the data.

\subsection{Synopsis}

Our development unfolds as follows. Section~\ref{sec:platonic} introduces
Platonic models as the true dynamics of a given physical system. This is what
models attempt to predict. Next, Section~\ref{sec:dynproc} formalizes partial
observations and the resulting stochastic processes over the observable degrees
of freedom, which we call dynamical processes. These are the main objects of
study. To set the stage for the development of implicit data-driven models,
Section~\ref{sec:explicit_models} first reviews the explicit physics-based
modeling approach. Next, Section~\ref{sec:physPO} gives the physics of partial
observations expressed in terms of Koopman and Perron-Frobenius operators. This
section also discusses connections to statistical mechanics and introduces
Maximum Entropy measures. 

Section~\ref{sec:Instant_Implicit_Models} overviews implicit data-driven models
and their Hilbert space formulation for the case of instantaneous prediction.
Section~\ref{sec:MZ} details the Mori-Zwanzig formalism, motivating
history-dependent models. Section~\ref{sec:delay_embeddings} discusses
histories of past observations in the form of delay-coordinate embeddings.
Section~\ref{sec:history-dep_models} then expresses the Mori-Zwanzig formalism
in terms of delay embeddings using Wiener projections. This provides the
formulation of history-dependent Hilbert space models, using both the
equilibrium invariant measure and nonequilibrium Maximum Caliber measures. In
the nonequilibrium case, the Maximum Caliber measures allow for the direct
construction of predictive distributions, providing insights into the
convergence behavior of optimal history-dependent models. 
Section~\ref{sec:examples} provides examples demonstrating the ability of data-driven models to implicitly learn the effects of the unobserved degrees of freedom.
Finally,
Section~\ref{sec:model_framework} uses the prior development to formally
connect implicit data-driven models with explicit physics-based models. This
shows the underlying similarity between the two approaches and offers a unified
implicit-explicit modeling framework. 

\section{Systems and Platonic Models}
\label{sec:platonic}

After centuries of intellectual inquiry, physical scientists collectively have
come to believe in having a solid grasp of the basic physics governing
measurable phenomena. For example, many Earth Science systems are governed by
classical field theories. Atmospheric circulation, shown in
Fig.~\ref{fig:modeling}, is governed by the laws of fluid mechanics and
thermodynamics \cite{ghil20a}.

Saying that one ``understands'' these system's basic physics means, more
specifically, that the governing principles are encapsulated in the form of
explicit differential equations-of-motion \cite{meiss07a}. Formally, the system
state $\omega$ evolves according to:
\begin{align*}
\dot{\omega} & = \frac{d \omega}{dt} \\
   & = \Phi(\omega)
  ~,
\end{align*}
where the governing equations $\Phi$ are a function of $\omega$. For
spatially-extended field theories, $\omega$ itself is a function of spatial
coordinates, too. $\Phi$ then typically includes finitely-many spatial
derivatives of $\omega$, signifying the state dynamics are governed by local
interactions.

The ``unreasonable effectiveness of mathematics'' in physics has been
repeatedly noted since Ref. \cite{wign60a} highlighted the puzzle. Noting
that governing equations $\Phi$ are almost always given in \emph{closed form}
the effectiveness is all the more intriguing.
Our development further highlights that demanding physical systems always be expressed in closed form rather restricts the class of mathematical models used to describe the physical world.

Here, we represent a given physical system as a differential dynamical system
$(\Omega, \Phi)$ that, for a shorthand, we call the \emph{Platonic model}. A
system's true dynamics, given by the Platonic model, may be well
approximated with closed-form equations of motion. The Navier-Stokes partial
differential equations come to mind as an approximation to the Platonic model
of fluid flow. However, we need not assume a particular functional form for
Platonic models.

That said, there are three important properties we do assume for Platonic models.
Note that we are primarily concerned here with phenomena that occur at
classical energy scales, such as found in Earth Systems. The first property is
that system states evolve continuously---they are continuous trajectories in
the state space over time.

The next two properties define what the system \emph{state} $\omega \in \Omega$ actually is. The second property assumes Platonic models are \emph{Markovian}:
Determining a later state $\omega_t = \Phi^t \omega_0$ only requires knowing
the state at a single prior time $\omega_0$. The third property assumes
Platonic models are \emph{deterministic}: The same initial condition $\omega_0$
always produces the same later state $\omega_t = \Phi^t \omega_0$.



The latter two properties impose a \emph{closure relationship} among the
\emph{degrees of freedom} constituting the system state $\omega$. That is,
$\omega$ is considered a vector with each component $\omega^i$ being a degree
of freedom. The dynamic $\Phi(\omega)$ captures the physically-relevant
interactions among the degrees of freedom by determining how they evolve
forward in time. The system's governing physics is appropriately captured or
modeled when, with sufficiently-many degrees of freedom comprising $\omega$,
there is a closure in their dynamics: For every $\omega^i$, its time evolution
is a deterministic and Markovian function of a subset of the other
$\{\omega^i\}$, i.e., the system state $\omega$.

As there are many parallels to statistical mechanics, note that there is an
important property we are \emph{not} assuming of $(\Omega, \Phi)$---that the
system is Hamiltonian. In the partially-observed setting, introduced shortly,
the Platonic model $(\Omega, \Phi)$ is analogous to a ``microsystem''.
Statistical mechanics would take it to be Hamiltonian. This is too restrictive
for our purposes. Importantly, Hamiltonian systems are conservative and
volume-preserving, via Liouville's theorem \cite{Wild98a}. 
Volume-preserving dynamics admit a natural invariant probability distribution over $\Omega$, known as the \emph{microcanonical ensemble} in statistical mechanics \cite{Wild98a}. While such invariant probability measures are convenient mathematically, many physical systems of interest display transient nonasymptotic behavior that cannot be captured by invariant measures. This is particularly notable for fluid flows.  

To accommodate nonasymptotic behaviors within our formalism, we do not assume
Platonic models are necessarily volume-preserving, although they may be. More
generally, while it is standard to assume the dynamics is
\emph{measure-preserving} such that there is a probability measure over
$\Omega$ which is invariant under $\Phi$, our formalism does not require an
invariant measure. Rather, one of our main contributions is introducing natural
time-dependent measures for partially-observed systems that can support
nonasymptotic behaviors. In the language of statistical mechanics, our approach
is a nonequilibrium formalism that generalizes the equilibrium setting
using asymptotic invariant measures. For more details on ergodicity,
invariant measures, and dissipative systems, see Appendix~\ref{app:ergodicity}.





Additionally, in what follows, we assume a system's dynamic is reversible,
so that:
\begin{align*}
    (\Phi^t)^{-1} = \Phi^{-t}
    ~.
\end{align*}
This, however, is an assumption for notional convenience and simplicity.
It can be lifted without much difficulty. An added advantage of our
time-dependent formulation is that we need not assume reversible dynamics. 
Note though that many systems of interest are reversible in this way, such as all finite-dimensional systems of ordinary differential equations. 

\section{Partial Observations and Dynamical Processes}
\label{sec:dynproc}

The semigroup formalism of dynamical systems \cite[Ch. 7]{laso94a} is particularly
apt for our development. Consider a dynamical system $(\Omega, \Sigma_\Omega,
\nu, \Phi)$. The state space $\Omega$ is a Euclidean space or manifold for
finite-dimensional systems or a general Hilbert space for spatially-extended
systems. $\Sigma_\Omega$ is the Borel $\sigma$-algebra and $\nu$ the Lebesgue
reference measure that gives a ``volume'' to state space.

$\Phi$ is the dynamic---the infinitesimal generator of a continuous semigroup
of measurable flow maps $\{\Phi^t : \Omega \rightarrow \Omega\}_{t \in
\mathbb{R}}$, with:
\begin{align*}
\Phi(\omega) & = \lim\limits_{\tau \rightarrow 0}
\frac{1}{\tau} \bigl(\Phi^{t+\tau}(\omega) - \Phi^{t}(\omega)\bigr) \\
  & = \frac{d}{dt} \Phi^t(\omega) |_{t=0}
  ~.
\end{align*}
Thus, the orbits $\{\omega_t =
\Phi^t(\omega_0) : t \in \mathbb{R}_{(\geq0)}\}$ are continuous functions of
time $t$. When the dynamic is specified by a system of differential equations,
$\Phi$ is the time derivative of the orbits:
\begin{align*}
\dot{\omega} & = \frac{d}{d t} \omega \\
   & = \Phi (\omega)
  ~.
\end{align*}


For a given dynamical system under study, let $x \in \mathcal{X}$ be the subset
of system variables that are observable, measurable, or generally accessible.
Through experimental or observational measurements or numerical simulations,
they may be collected in a \emph{time series} $\{x_0, x_1, \dots,
x_{T-1}\}$---a time-ordered set of observations of $x$ taken at uniform time
intervals $\{t_0, t_1, \dots, t_{T-1}\}$ with $t_i = (i-1)\Delta t$. The
observations $x$ are generated by the dynamical system under the continuous and
measurable mapping $X \; : \; \Omega \rightarrow \mathcal{X}$ so that $x_t =
X(\omega_t)$. In practice, the measurement observables are given as a vector of
real numbers, so that $\mathcal{X} = \mathbb{R}^n$.

We are interested in the case of a \emph{partially-observed} dynamical system
for which the map $X$ is many-to-one and not invertible. Due to this, an
observation $x_t$ is insufficient for determining the full state $\omega_t$ of
the underlying dynamical system at any given time. That is, there are unobservable,
unmeasurable, or inaccessible degrees of freedom in $\omega$. And so,
measurement data can only ever provide a limited view of the system's true
state $\omega$. An important example is weather prediction, shown in
Fig.~\ref{fig:modeling}.

We refer to collections of arbitrarily-long time series of observables
$\dynproc$ as a \emph{dynamical process}, signifying that it is a stochastic process derived from a deterministic dynamical system through partial observations. They are the objects we wish to
model. If the underlying system is governed by noninvertible dynamics we
consider the time index of a dynamical process to correspond to
\emph{observation time}. That is, $x_0$ is not an initial condition, but rather
the present moment of observation. The leading dots then indicate that we allow
measurements from arbitrarily far in the past.

Various properties of dynamical processes will be given shortly, using the Koopman and Perron-Frobenius operators. First though, we detail the standard approach for modeling partially-observed systems using physics-based models.

\section{Explicit Predictive Models}
\label{sec:explicit_models}

\begin{figure*}
\begin{center}
\includegraphics[width=1.0 \textwidth]{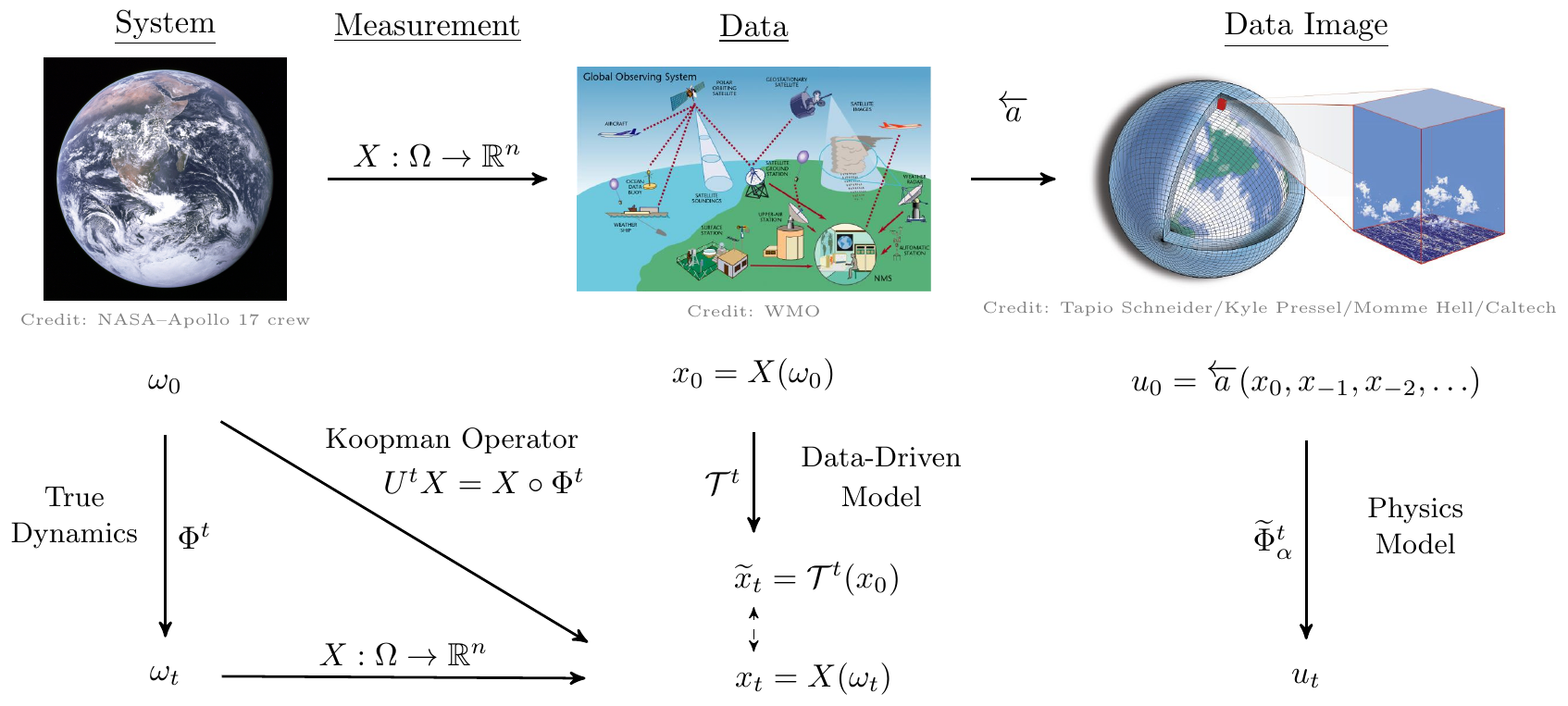}
\end{center}
\caption{Predicting complex systems (left) from partial observations:
	Instantaneous data-driven modeling (middle) versus physics-based models
	(right). Instrument measurements provide a partial view $x$ of the true
	system state $\omega$ through a noninvertible mapping $X$. From an initial
	measurement observation $x_0 = X(\omega_0)$, the value of the instruments
	at later time $t$ is found by letting the actual system evolve, given by
	the dynamic $\Phi^t$, and taking a measurement $x_t = X(\omega_t)$ at this
	time. Koopman operators $U^t$, provide an alternate point of view by
	providing a \emph{future measurement} function $x_t = U^t X(\omega_0)$ that
	gives the instrument readings $x_t$ at time $t$ from the current system
	state $\omega_0$. It provides the ground-truth that data-driven models
	$\mathcal{T}^t$ try to approximate as functions from current observation
	$x_0$ to future observation $x_t$. In contrast, physics-based models create
	a coarse-grained approximation $u_0$ of the full system state $\omega_0$
	most consistent with past observations using data assimilation and model inversion as $u_0 = \protect\overleftarrow{a}(x_0, x_{-1}, x_{-2}, \ldots)$. Data images are then evolved through
	numerical approximation $u_t = \widetilde{\Phi}^t(u_0)$ of the system
	dynamics $\omega_t = \Phi^t(\omega_0)$. 
	}
\label{fig:modeling}
\end{figure*}

Given a physical system's Platonic model---its governing physics---and partial
observations from instrument measurements, how do we predict the system's
future behavior? Our main interest is to explain the effectiveness of
\emph{implicit} approaches learned by data-driven models. To set the stage,
though, we first overview the more familiar \emph{explicit} approach using
physics-based models. Figure~\ref{fig:modeling} shows the relation between
data-driven and physics-based methods for modeling systems from partial
observations. After formulating the physics underlying implicit data-driven
models, a formal connection with explicit physics models is given in
Section~\ref{sec:model_framework}.

\subsection{Physics-Based Models}

In essence, physics-based models simply attempt to solve the governing
equations that constitute or well-approximate the Platonic model $\Phi$. There
are three main challenges when predicting a physical system using differential
equation models: nonlinearity, high-dimensionality, and calibration.

First, most systems of interest are governed by nonlinear equations that cannot
be solved analytically. Thus, numerical approximations to solutions are
necessary. For complicated systems like those encountered in Earth
Sciences, this introduces a second challenge.

Even with the arrival of massive high performance computing, today's largest
machines still do not have the computational resources required to fully
account for all known physical effects in a system at the necessary scales.
This second challenge is certainly the case for numerical models of the
atmosphere, as depicted in Fig.~\ref{fig:modeling}'s right column. The effects
that are not directly computed are accounted for using \emph{parametrization}
schemes to replace processes that are too small-scale or complicated to be
directly computed in the model. This simplifying procedure produces a
deterministic, Markovian closure. 

While parametrization schemes are often heuristic choices, increasingly they
are being informed by separate models specifically targeting the effects being
parameterized. This includes, for example, using cloud-resolving models to
inform cloud formation parametrizations in large-scale atmospheric models.
Moreover, for spatially-extended field theories, continuous spatial coordinates
must be discretized into a finite grid or mesh and then the effects of
subgrid-scale processes must be parameterized using approximate closure models.
However they are arrived at, the parametrizations represent a modeler's
choices, and these choices necessarily induce \emph{conceptual error} that
affects the model's predictive capabilities. Poor choice of generic model may
also lead to conceptual error in prediction. 


Assume, for a given physical system, that we know effective differential
equations $\Phi(\omega)$ that govern the system. A \emph{generic model} of
$\Phi(\omega)$ is an auxiliary set of equations $\widetilde{\Phi}^\tau_\alpha:
u_t \mapsto u_{t+\tau}$ whose solutions $\{u_t\}$ can be solved numerically
and that approximate the solutions of $\Phi$. The generic model typically
contains a set of parameters $\alpha$ that include those associated with
parametrization schemes as well as physical parameters of the model, such as
viscosity in the Navier-Stokes equations. The generic model
$\widetilde{\Phi}^\tau_\alpha$ acts on \emph{data images} $u_t$ that are
coarse-grained approximations of the state $\omega_t$ of the physical system.
Neglecting numerical round-off error, numerical models are also Markov and
deterministic, like the differential equation models they approximate.

\subsection{Data Assimilation}

The final challenge in using a generic physics model to predict the future
behavior of a partially-observed physical system comes during
\emph{calibration}. The generic model must be made into a \emph{specific model}
that appropriately captures the particular circumstances of the physical system
of interest. This includes specification of the parameters $\alpha$ and
boundary conditions, as well as \emph{initialization} of the model. (For
simplicity, we include specification of the boundary conditions in $\alpha$.)
The Markov property allows for generating an orbit $u_t$ of the specific model from
a single initial state $u_0$. For this orbit to provide a prediction of the
true system's orbit $\{\omega_t\}$ requires aligning the model's initial state
$u_0$ as well as possible to the true physical system's initial state
$\omega_0$.  

To emphasize the difficulty of initialization in particular, consider the
commonly-encountered case of predicting a spatially-extended system using
approximated solutions of a classical field theory---i.e., $\Phi$ is a set of
partial differential equations. It is not possible to determine the system's
configuration over a continuum of spatial coordinates. Rather, as depicted in
the top of Fig.~\ref{fig:modeling}'s middle column, measurements derive from a
variety of instruments collecting data over a relatively small subset of the
spatial domain. However, solving the model equations---say, using a finite
element method---requires an initial condition on a grid over the spatial
domain. Our instruments, though, do not necessarily provide full coverage over
the grid. Thus, the calibration methods produce a data image $u_0$ (top right
of Fig.~\ref{fig:modeling}) that represents inferred values over the full grid.

Model calibration, including parameter and boundary condition specification, as
well as initialization of the data image $u_0$, are carried out using
\emph{model inversion} and \emph{data assimilation} \cite{bout99a,sanz18a}.
Since these techniques require multiple past observations, calibration is
sometimes also referred to as \emph{history matching}. Given a history of past
observations $\past_t^k := \{x_t, x_{t-1}, \ldots, x_{t-k}\}$, calibration
attempts to find the initial data image $u_t =
\overleftarrow{a}(\overleftarrow{x}_t^k)$ and parameter set $\alpha$ such that
the model output $\{u_t, \widetilde{\Phi}^{-1}_\alpha(u_t), \ldots,
\widetilde{\Phi}^{-k}_\alpha(u_t)\}$ is as consistent with the past
observations $\past_t^k$ as possible. (Recall that we are assuming reversible
dynamics for notational simplicity, but this is not strictly required.)

Due to the many-to-one nature of the partial observation map $X$, the
calibration process is typically not unique. Multiple parameter sets and
initial data images may produce orbits of data images that are equally
consistent with the observations up to time $t$. Therefore, there will be
multiple specific models that are equally consistent with past observations,
but make different predictions for future behaviors. Therefore, a specific
model used for prediction generally has \emph{calibration error}. This combined
with conceptual error leads to the model's overall prediction error. Prediction
error can accumulate rapidly, particularly for deterministic chaotic systems
whose inherent instabilities exponentially amplify small variations. This is
one reason why weather is so hard to predict.

We stress here that model parametrizations and initialization $u_0 = a(x_0)$
are both means of \emph{explicitly} accounting for unobserved or unrepresented
degrees of freedom not in $x=X(\omega)$. For instance, in atmospheric
circulation, imagine we do not have instruments on a remote island in the
Pacific. As a consequence, atmospheric variables---temperature, pressure, wind
speed, and the like---at that spatial location are not in $x=X(\omega)$.
However, when a data image $u$ is created these variables \emph{are}
approximated at that location.

Note also that the primary concerns are the predictive capability of physics
models and how it relates to the Platonic model's true dynamics. In a sense,
though, we are agnostic as to whether a specific model is \emph{valid} or not
\cite{koni92a,carr93a}. Loosely speaking, validity measures how well a physical
system's specific model approximates its Platonic model. The specific model's
predictive skill is, of course, related to how well it approximates the
Platonic model. And, this is a question we care about here.

That said, there is a deeper concern about how well a specific model
approximates the Platonic model. This involves the question of how much we can
infer about the underlying physical and causal processes governing the true
system, given a specific model of that system and its predictive capability.
That is, how well can we explicitly formulate a Platonic model given a skillful
specific model? It is in this deeper mechanistic sense that we are agnostic to
the question of model validity. As the saying goes, ``All models are wrong, but
some are useful''~\cite{box76a}.



\section{Physics of Partial Observations}
\label{sec:physPO}

Platonic equation-of-motion models are given in terms of the underlying system
state---the full set of degrees of freedom. Due to their explicit nature, the
connection between physics models and the system's true governing physics
described by Platonic models is clear. Data-driven models of partially-observed
systems do not generally attempt to explicitly infer the full Platonic system state, as
physics-based models do. And so, it is less clear how they relate to the
physics of Platonic models. To discuss Platonic models and the true governing
physics in a meaningful way for partially-observed system requires the
operator-theoretic formulation of dynamical systems \cite{laso94a,berr20a}.
Both Koopman and Perron-Frobenius operators, defined shortly, provide
alternative descriptions of a system's temporal evolution: Koopman operators
give the evolution of observables, while Perron-Frobenius operators evolve
state distributions. These \emph{evolution operators} are the classical analogs
of the Heisenberg and Schr\"odinger formulations of quantum mechanics,
respectively.

\subsection{Koopman Operators}

A \emph{Koopman operator} $U^t$ acts on functions of the system state, known as
\emph{observables} $f : \Omega \rightarrow \mathbb{R}$, where $f$ is an element
of a function space $\mathcal{F}$. The action of $U^t : \mathcal{F} \rightarrow
\mathcal{F}$ on observable $f$ is given by composition with the dynamic
$\Phi^t$, also known as the \emph{pullback} of $f$ along $\Phi^t$:
\begin{align}
U^t f & = f \circ \Phi^t, \\
[U^t f](\omega) & = f_t(\omega) \nonumber \\
  & :=  f\bigl(\Phi^t(\omega) \bigr)
  ~.
\label{eq:koopman}
\end{align}
That is, $U^t$'s action on observable $f \in \mathcal{F}$ gives the
time-shifted observable $f_t = U^t f$ whose value at state $\omega$ is obtained
by evaluating $f$ at the future state $\omega_t = \Phi^t(\omega)$. Recall that
the flow maps $\{\Phi^t\}$ form a semigroup in that $\Phi^{t+s} = \Phi^t \circ
\Phi^s$. The set $\{U^t\}$ inherits this semigroup structure, so that $U^t
\circ U^{\Delta t} = U^{t + \Delta t}$.


Each $U^t$ is a linear infinite-dimensional operator when $\mathcal{F}$ is a
vector space. As discussed more below in relation to Perron-Frobenius
operators, it is most natural to take $\mathcal{F} = L^\infty(\Omega,
\nu)$---the bounded functions of $\omega$---but the square-integrable functions
$L^2(\Omega, \nu)$ are often used for mathematical convenience. The following
uses Koopman operators on $L^2(\Omega, \nu)$ since it is a
Hilbert space and the development requires orthogonal projections.

Recall that we are interested in observable functions $X$ that are generally
multidimensional. With this, an observable function $f$ is a component of a
vector-valued observable function; e.g., $f = X_i$. A Koopman operator that
acts on an observable $X$ is then the product over the component operators
acting on $X_i$. To avoid excessive notation, we denote these product operators
as $U^t$.

For a dynamical system with initial condition $\omega_0$ at time $t_0$, the
measurement observable at a later time $t > t_0$ is given by:
\begin{align*}
x_t & = X(\omega_t) \\
  & = X\bigl(\Phi^t(\omega_0)\bigr) \\
  & = X_t(\omega_0) \\
  & = [U^t X] (\omega_0)
  ~.
\end{align*}
The dynamical process, therefore, is a function of the underlying system's (unknown) initial
condition:
\begin{align}
\dynproc & = \nonumber \\
  \{ \ldots, U^{-1}X(\omega_0) &, U^0 X(\omega_0), U^1X(\omega_0), \dots \}
  ~.
\label{eqn:dynproc}
\end{align}
This transparently relates the evolution of partial measurement observations
of the dynamical process $\dynproc$ to the physics of the Platonic model
$\Phi^t(\omega_0)$ through the action of Koopman operators on the observable
map $X$.

\subsection{Perron-Frobenius Operators}

Koopman operators connect dynamical processes to the Platonic model $(\Omega,
\Phi)$ via an unknown initial Platonic state $\omega_0$. If we do not seek to
directly infer $\omega_0$, as done with physics models, it becomes useful to
formulate the problem in terms of distributions over possible $\omega_0$. The
dynamics of these distributions is provided by \emph{Perron-Frobenius
operators}.

In appropriately defined spaces, Perron-Frobenius operators are dual to Koopman
operators. It is most common to consider Perron-Frobenius operators acting on
$L^1(\Omega, \nu)$ densities and, thus, their Koopman duals evolve observables
in $L^\infty(\Omega, \nu)$. However, as often done, the following considers both
operators acting on $L^2(\Omega, \nu)$ functions. In this case,
Perron-Frobenius operators act on $L^2$ measures. If the $L^2$ measure
$\nu_\rho$ is absolutely continuous with respect to the reference measure
$\nu$, $\nu_\rho$ is related to the density $\rho$ through the reference
measure:
\begin{align*}
\nu_\rho(B) = \int_B \rho d\nu
  ~,
\end{align*}
for density $\rho \in L^1(\Omega, \nu)$ and $B \in \Sigma_\Omega$.

For continuous-time dynamical systems there is a continuous
semigroup $\{P^t\}$ of Perron-Frobenius operators that evolve measures $\mu$
through the \emph{pushforward} of $\mu$ along $\Phi^t$:
\begin{align}
\mu_{t} & = P^t \mu  \nonumber\\
  & := \mu \circ \Phi^{-t}
  ~.
\label{eq:perronfrobenius}
\end{align}


The measure $\mu_t$ defines the probability space $(\Omega, \Sigma_\Omega,
\mu_t)$ that quantifies uncertainty in system state $\omega_t$ at time $t$.
In turn, this casts observables, given by the measurable map $X: \Omega
\rightarrow \mathcal{X}$, as random variables $X_t$ distributed according to
the pushforward measure:
\begin{align*}
\mu_t^X (B_{\mathcal{X}}) = \mu_t \left(X^{-1} (B_{\mathcal{X}}) \right)
  ~,
\end{align*}
for $B_{\mathcal{X}} \in \Sigma_{\mathcal{X}}$. Thus,
we can write $X_t$'s distribution in terms of the initial measure $\mu_0$:
\begin{align}
\Pr(X_t \in B_{\mathcal{X}}) & = \int_{B_\mathcal{X}} d\mu_t^X \nonumber \\
  & = \int_{X^{-1}(B_{\mathcal{X}})} d\mu_t \nonumber \\
  & = \int_{\Phi^{-t}\bigl(X^{-1}(B_{\mathcal{X}})\bigr)} d\mu_0
~.
\label{eq:temporal_measures}
\end{align}

This is analogous to writing, as done in Eq.~(\ref{eqn:dynproc}), observations
$x_t$ in terms of the initial state $\omega_0$ and the action of Koopman
operators on the measurement observable $X$. Recall that the two operators are
dual, so that these two perspectives are equivalent. If there is initial
uncertainty over system states, then the observables become random variables.
The Koopman operator then evolves observable random variables that are
distributed according to the action of
Perron-Frobenius operators on the initial distribution.

Thus, given an initial uncertainty measure over system states, a dynamical process is a \emph{stochastic process} $\stocproc$---a
time series of random variables---with \emph{realizations} $\dynproc$. Note
that the random variables in $\stocproc$ are actually (measurable) functions of
two variables: $X_t = X(t, \omega_0)$. Fixing $\omega_0$ produces a
realization, or \emph{sample path}, $\realization$ of the stochastic process.
From our setup, the realization $\realization$ for a given $\omega_0$ is the
result of applying the map $X$ to each $\omega_t$ in the orbit generated by
$\omega_0$. We consider continuous maps $X$ so that realizations $\{x_t =
X(\omega_t) : \omega_t = \Phi^t(\omega_0)\}$ are also continuous curves in $t$.

We emphasize again that evolution operators are defined in Eqs.
(\ref{eq:koopman}) and (\ref{eq:perronfrobenius}) in terms of the Platonic
model $\Phi$. As such, they are yet other ways of expressing the true governing
physics of a given system. In particular, they provide the true physics of
partial observations.


\subsection{Nonequilibrium Statistical Mechanics}

Equation~(\ref{eqn:dynproc}) expresses the time series of measurement
observations in terms of Koopman operators and an unknown initial Platonic
state $\omega_0$. In contrast, Eq.~(\ref{eq:temporal_measures}) expresses the
measurement observables as a continuous stochastic process using
Perron-Frobenius operators and an initial probability distribution $\mu_0$ over
the Platonic states. To compensate for not knowing the exact initial state
$\omega_0$, one can ask, is there a natural choice for an initial distribution
$\mu_0$ over $\Omega$ induced by observations? This key question leads directly
to statistical mechanics.

The standard choice for $\mu_0$ is the invariant measure $\mu_*$ given
by $P^t \mu_* = \mu_*$. For the ergodic systems considered here $\mu_*$ is
guaranteed to exist and to be reached asymptotically (see Appendix
\ref{app:ergodicity}). The following employs this commonly-invoked
``equilibrium case'' to review instantaneous data-driven models. Note that, by
definition, taking the invariant measure $\mu_*$ as $\mu_0$ leads to $\mu_t =
\mu_*$ for all $t$. Due to this, the random variable observables in Eq.
(\ref{eq:temporal_measures}) have time-independent distributions. In this case,
the stochastic process over measurement observables is a \emph{stationary}
stochastic process. Clearly though, assuming the invariant measure $\mu_*$
precludes nonasymptotic ``nonequilibrium''behaviors that we ultimately wish to
also capture.

The preceding defined dynamical processes as stochastic processes generated by
deterministic dynamical systems. To set the stage for the nonequilibrium
generalization with time-dependent measures used later for history-dependent
models, recall that underlying system states $\omega_t$ can not be uniquely identified from an observation $x_t$
due to the noninvertibility of the measurement observable function $X$. This
setup admits a natural nonasymptotic measure induced by a single observation
$x_t = X(\omega_t)$ that we now define.

Consider a dynamical system $(\Omega, \Sigma_\Omega, \nu, \Phi)$ and a single
observation $x_t = X(\omega_t)$ at an arbitrary time $t$. Since the observation
mapping $X$ is not invertible there can be many $\omega_t \in \Omega$ yielding
the observed value $x_t$ under $X$. (This is directly related to the non-uniqueness of model inversion when assimilating physics-based models). 
Thus, for a given observation $x_t$ define
the set $B_t \in \Sigma_\Omega$ as:
\begin{align}
    B_t = X^{-1}(x_t) =  \{\omega_t \in \Omega \; | \; X(\omega_t) = x_t\}
    ~.
    \label{eq:instantaneous_B}
\end{align}
Note that $B_t$ is $\nu$-measurable. 

Following Refs. \cite{Jayn57a,Gran08a}'s minimal bias argument there is a
natural measure $d\mu_t = \rho_t d\nu$ defined through the density $\rho_t$
that is constant over $B_t$ and zero elsewhere, so that $\Pr(\omega_t \in b
\subseteq B_t) = \nu(b) / \nu(B_t)$. The \emph{Maximum Entropy Principle} (MEP)
says that the distribution which maximizes entropy subject to known constraints
creates the minimally-biased prior distribution that is spread out as much as
possible, up to given constraints. If the only constraint given is the support
set, MEP reduces to the \emph{Principle of Indifference} and assigns uniform
probability over the set.

In what way is the noninvariant measure $\mu_t$ a nonequilibrium generalization
of the equilibrium measure $\mu_*$? The nonequilibrium behaviors allowed by
ergodic systems with dynamics $\Phi$ which have no explicit time-dependence are
those of relaxation processes \cite{mack92a}. According to the attractor-basin
formalism described in Appendix \ref{app:ergodicity}, these processes limit to
equilibrium distributions given by the invariant measure $\mu_*$. Theorem 4.5
in Ref.~\cite{mack92a} establishes the correspondence between the invariant
measure $\mu_*$ and thermodynamic equilibrium for these systems. Hence, any
other measure $\mu$ is a nonequilibrium distribution that asymptotically limits
to the equilibrium distribution. The measure $\mu_t$ is a nonequilibrium
measure naturally induced through partial observations. 

Note that there is a wide range of nonequilibrium phenomena beyond relaxation
processes. For instance, an invariant measure may correspond to an equilibrium
steady state or a nonequilibrium steady state \cite{oono98a} that absorbs and
dissipates energy from its surroundings. These more general
far-from-equilibrium processes that include thermal driving require explicit
time-dependence in the dynamics \cite{te19a}. Detailed thermodynamic analysis
is not our primary concern as yet, but the formalism introduced here readily
extends to such settings by including explicit time-dependence in the Koopman
and Perron-Frobenius operators. See, for example, the dynamics governed by
Ref.~\cite{sema22a}'s time-dependent rate-matrices. In that language, the
development here applies in the special case of a fixed time-independent
protocol with relaxation to the associated invariant distribution.


As seen shortly, the two measures $\mu_*$ and $\mu_t$ represent different sets
of assumptions used to motivate and interpret the behavior of data-driven
models. Neither is typically known explicitly, but is rather inferred
approximately from observations. Kernel methods are particularly useful for
this in practice \cite{berr20a,brod20a}. The nonequilibrium measure $\mu_t$ is
more closely aligned with the modeling approach of physics-based models. Its
construction requires knowledge of the set $B_t$ of Platonic states consistent
with the observation $x_t$, much like the construction of the data image $u_t$
that is the approximation of the Platonic state most consistent with $x_t$.
Ultimately though, both $\mu_*$ and $\mu_t$ are insightful in their own way for
understanding implicit data-driven models, and so both are discussed in
detail in what follows.

\section{Instantaneous Implicit Models}
\label{sec:Instant_Implicit_Models}

With the physics of dynamical processes laid out using the machinery of Koopman
and Perron-Frobenius evolution operators, we return to the question of optimal
prediction. Section ~\ref{sec:explicit_models} outlined the challenges of using
physics-based modeling for prediction. Circumventing explicit inference of
unobserved degrees of freedom requires learning, directly from observation,
evolution rules for the variables that are accessible through instrument
measurements. Pushing this further, we explore learning \emph{implicit models}
that predict the evolution of the observables---models given in a more
flexible, possibly more abstract, form than differential equations-of-motion.

\subsection{Instantaneous Predictive Distributions}

The most basic form of prediction for a dynamical process is instantaneous:
Given a single observation $x_t = X(\omega_t)$, predict the observable at a
single time in the future $x_{t+\tau} = X(\omega_{t+\tau})$. Before reviewing
the functional Hilbert space approach for learning instantaneous implicit
models, we first theoretically analyze the problem using evolution operators. We
argue that the maximal instantaneous predictive information available is given
in an instantaneous predictive distribution. We will later see that the optimal
Hilbert space model for instantaneous prediction is the expectation value of
the instantaneous predictive distribution.

Given a single observation $x_t$, Eq. (\ref{eq:instantaneous_B}) defined $B_t$
as the set of all possible Platonic states $\omega_t$ consistent with the
observation $x_t$ such that $X(\omega_t) = x_t$. This then defines the set of
all possible observables $x_{t+\tau}$ that may be seen at a later time by
evolving each $\omega_t \in B_t$ under $\Phi^\tau$ and applying the observable
mapping $X$. Said another way, the set of all possible future observables
$x_{t+\tau}$ is given through the action of the Koopman operator by applying
the time-shifted observable $X_\tau = [U^\tau X](\omega_t)$ to all $\omega_t$
in $B_t$.

Furthermore, we use the MEP measure $\mu_t$ over $B_t$ and the Perron-Frobenius
operator to define the distribution over possible future observables, supported
on the set $\{x_{t+\tau} = [U^\tau X](\omega_t), \; ~\mathrm{for~all}~ \omega_t
\in B_t\}$. This distribution---the \emph{instantaneous predictive
distribution}---is given as the pushforward of the time-evolved measure
$\mu_{t+\tau} = P^\tau \mu_t$ along $X$, following Eq. (\ref{eq:temporal_measures}):
\begin{align}
    \Pr(U^\tau X | X_t = x_t) \; \sim \; \mu_{t+\tau}^X
    ~.
    \label{eq:instant_pred_dist}
\end{align}
To define the instantaneous predictive distribution, we need \emph{some}
initial measure $\mu_t$. Without additional information on the system, the
choice of a MEP measure is most natural. What matters for our purposes is that
$\mu_t$ is supported on the set $B_t$ and, thus, the instantaneous predictive
distributions are supported on $\{x_{t+\tau} = [U^\tau X](\omega_t), \;
~\mathrm{for~all}~ \omega_t \in B_t\}$. In practice, these measures are
estimated empirically from data and the MEP is not typically invoked for
$\mu_t$'s empirical construction.

Also note that the formalism for instantaneous models we now review is given in
terms of the equilibrium measure $\mu_*$, as is standard. 
However, instantaneous predictive
distributions cannot be expressed directly in terms of $\mu_*$. That said,
the nonequilibrium construction of predictive distributions just given
is instructive for understanding instantaneous models built using $\mu_*$. 
The equilibrium and nonequilibrium formulations of data-driven models are more
closely connected below for history-dependent models using Wiener projections. 
Our introduction of nonequilibrium measures is most impactful for history-dependent models, as they provide insights not available through use of the invariant measure.

\subsection{Instantaneous Data-Driven Models}

We now review instantaneous data-driven models and their Hilbert space
formalism. Following established practice, we use the invariant measure $\mu_*$.

Given the current observation data $x_t$, the goal is to construct a
model $\iddmodel: \mathcal{X} \rightarrow \mathcal{X}$, called a \emph{target
function}, that predicts what the instruments will read at a later time
$t+\tau$ \cite{alex20a}. This is depicted in Figure~\ref{fig:modeling}'s middle
column. On the one hand, recall that for physics-based models we assume the
system's governing equations of motion $\Phi$ are known, and that one of the
main challenges is to infer from partial observations $x_t = X(\omega_t)$ the
underlying Platonic state $\omega_t$ that the equations evolve. On the other
hand, the data-driven paradigm flips this around to work directly with the
measurement observations $x_t$, without directly inferring $\omega_t$. In point
of fact, an appropriate set of governing equations for $x_t$ is generally not
know a priori. They may not even be desired. Instead, the goal is to
\emph{learn} a model $\iddmodel$ from the measurement data.

In some cases we can learn $\iddmodel$ as closed-form equations using Galerkin
projections of $\Phi$ onto $\mathcal{X}$ \cite{rowl04a}. In many cases, though,
the evolution of $x_t$ cannot be adequately described by a set of closed-form
equations \cite{Crut87a}. Thus, we seek more general forms for
$\mathcal{T}^\tau$ that are measurable mappings from $\mathcal{X}$ into
$\mathcal{X}$. For example, neural networks \cite{chat20a} are universal
function approximators \cite{rack20a} and so are able, in principle, to
represent $\mathcal{T}^\tau$.

As the name suggests, a modeler cannot simply write down an implicit model.
Rather, implicit models are implemented algorithmically and learned from
data. From the discussion above, the Koopman operator provides the ground truth
for prediction---the equivalent of the Platonic model. Following
Ref.~\cite{alex20a}, the mean-squared error for model $\iddmodel$ is given as:
\begin{align}
\big\|\iddmodel \circ X - U^\tau X \big\|^2_{L^2(\mu_*)}
  ~.
\label{eq:iddmodel}
\end{align}

From the Koopman operator definition and Fig.~\ref{fig:modeling}'s commutation
relations, the measurement data $\timeseries$, used to learn $\iddmodel$,
contain samples of the Koopman operator's action. That is, for an observation
$x_t = X(\omega_t)$, a later observation is:
\begin{align*}
x_{t+\tau} & = X(\omega_{t+\tau}) \\
  & = X_\tau(\omega_t) \\
  & = [U^\tau X](\omega_t)
  ~.
\end{align*}

Empirically, the Koopman operator's action is approximated through the action
of the \emph{shift operator} \cite{alex20a,berr20a}. And so, the ground truth
for training $\iddmodel$ is found by simply looking up in the observed data
$\timeseries$ what happens after time $\tau$. In this way, given $x_t \in
\timeseries$, $0 \leq t < (T-\tau)$, $\widetilde{x}_t = \iddmodel(x_t)$ is the
prediction made by $\iddmodel$. With the ground truth given by $x_{t+\tau} \in
\timeseries$, a parametric model (e.g., neural network) $\iddmodel$ is trained
by minimizing $\|x_{t+\tau} - \widetilde{x}_t\|^2$ over the training data, $0
\leq t < (T-\tau)$.

\subsection{Analog Forecasting}

Analog forecasting, dating back at least to Ref. \cite{lore69a}, is one of the
oldest methods for approximating $\iddmodel$ implicitly from data. The basic
procedure is to predict a system's future by finding the value recorded in past
observations (the analog) that is most similar to the present observation and
then use the following value in the recorded history as the forecast.

Formally, let $\{x_0, x_1, x_2, \ldots, x_T\}$ be the finite set of historical
observations---the training data set. Then, given the current observation $X_t
= x$, with $t > T$, identify $x$'s analog $x_a$ in the training set. This is
typically implemented with Euclidean distance:
\begin{align*}
a = \argmin\limits_{i \in \{0, \ldots, T-\tau\}} D(x, x_i)
~.
\end{align*}
The forecast $x_{t+\tau}$ of $x$ for $\tau$ time steps into the future is
given by the analog forecast $x_{a+\tau} \in \{x_0, x_1, x_2, \ldots, x_T\}$.
That is, the analog forecast simply looks up what happened in the training data
set $\tau$ time steps after the analog:
\begin{align}
    \iddmodel_{\text{AF}}(x) = x_{a+\tau}
    ~.
\label{eq:analog_forecast}
\end{align}
Analog forecasting is used for the data-driven prediction examples given below in Section~\ref{sec:examples}. 

\subsection{Optimal Hilbert Space Models}

As data-driven models, target functions $\iddmodel$ map a single input to a
single output. 
As such, target functions live in a
function space. Since inner products and orthogonal projections play an
important role in the development, we seek target functions as elements of a
Hilbert space. The following reviews the Hilbert space formulation of
instantaneous data-driven models \cite{alex20a,berr20a,gila21a}.

Recall that partial observations of a dynamical system $(\Omega, \Phi)$ induce
a time-dependent probability measure $\mu_t$ over $\Omega$. For simplicity
when using instantaneous models, in the asymptotic limit we employ the invariant
ergodic measure $\mu_*$. This leads to the probability space $(\Omega,
\Sigma_\Omega, \mu_*)$ and measurement observables as random variables given by
the measurable map $X: \Omega \rightarrow \mathcal{X}$.  The space
$\mathcal{X}$ is often referred to as the \emph{covariate space} and $X$ the
\emph{covariate map}.

More generally, we may consider a \emph{response space} $\mathcal{Y}$ and the
(measurable) \emph{response map} $Y: \Omega \rightarrow \mathcal{Y}$. The
target function is then a measurable mapping from covariates to a response:
$\iddmodel: \mathcal{X} \rightarrow \mathcal{Y}$. In our dynamical setting, the
covariate and response spaces are the same---the observation instrument
readings: $\mathcal{Y} = \mathcal{X} = \mathbb{R}^n$. The response map is
the time-shifted measurement observable $Y = X_\tau = X \circ \Phi^\tau: \Omega
\rightarrow \mathcal{X}$. Note that $X$ and $Y = X_\tau$ are both random
variables over the same probability space $(\Omega, \Sigma_\Omega, \mu_*)$
since $x_t = X(\omega_t)$ and $y_t = x_{t+\tau} = X_\tau(\omega_t)$. This is
what makes $\iddmodel$ predictive.

Consider the Hilbert spaces:
\begin{align}
\label{eq:HilbertSpaces}
H & := \{f: \Omega \rightarrow \mathcal{X} :
  \int_\Omega \|f^2(\omega)\|^2 d\mu_*(\omega) < \infty\}
  ~, \\
V & := \{g: \mathcal{X} \rightarrow \mathcal{X} : g \circ X \in H\}
  ~,~\mathrm{and} \nonumber \\
H_X & := \{f \in H : f = g \circ X ~\mathrm{for~some}~ g \in V\}
  ~. \nonumber
\end{align}
Note that $H = L^2(\mu_*)$---the set of square-integrable functions \emph{of
the full system state $\omega$}. And, $H_X$ is the Hilbert subspace of
$L^2(\mu_*)$ containing functions $f_X$ that depend \emph{only} on the
observable degrees of freedom $x_t = X(\omega_t)$. Then $V = L^2(\mu_*^X)$,
where $\mu_*^X$ is the pushforward of $\mu_*$ along $X$, is the set of
functions over the measurement observables such that composition with the
observable map $X$ is square integrable. Since the observables $X \in H$ are
what is accessible, the set $V$ is what is at our disposal to build implicit
models. However, since Koopman operators $U^\tau$ act on functions in $H$, we must
compose elements of $V$ with $X$ for proper comparison with $U^t$'s action.
Specifically, the ground-truth for the future observation is given by $X_{t+\tau} = U^\tau X$ which lives in the space $H$, while the functions available for us to learn are in the subspace $H_X$.

Equation~(\ref{eq:iddmodel}) identifies the unique minimizer---the optimal
$\iddmodel$. Denote it $Z^\tau$. In statistics, this estimator is known as the
\emph{regression function}, as well as the \emph{conditional expectation
function}. That is:
\begin{align}
\mathbb{E}[U^\tau X | X] & = Z^\tau \nonumber \\
  & := \argmin\limits_{g \in V} \|g \circ X - U^\tau X\|^2_{L^2(\mu_*)}
  ~.
\label{eq:target_estimator}
\end{align}
The conditional expectation $\mathbb{E}[\cdot | X]$ is the (nonlinear)
orthogonal projection $P_X: H \rightarrow H_X$ from $H$ onto $H_X$. Thus,
$Z^\tau = P_X U^\tau X_t$.
This is the best approximation of $X_{t+\tau} = U^\tau X \in H$ available using functions restricted to the subspace $H_X$.


For a given learned target function (data-driven model) $\iddmodel$, we
decompose its error via the regression function $Z^\tau$:
\begin{align}
\mathcal{E}(\iddmodel) & := \|\iddmodel \circ X - U^\tau X\|^2 \\
  & = \Theta(\iddmodel) + \Xi^\tau_X
  ~.
\end{align}
The \emph{excess generalization error} $\Theta(\iddmodel) = \|\iddmodel -
Z^\tau\|^2$ measures how far a given model is from the optimal solution, while
$\Xi^\tau_X$ is the \emph{intrinsic error} due to the partial observations $X$
of a given system $(\Omega, \Phi)$. For a given physical system with instrument
measurements $X$, the regression function $Z^\tau$ represents the maximum
predictive skill an instantaneous data-driven model can achieve.  $\Xi^\tau_X$
is then the unavoidable error incurred from only being able to measure $X$.
Increasing instrument coverage and expanding $X$ can decrease $\Xi^\tau_X$.

The conditional expectation $\mathbb{E}[U^\tau X | X]$ can be expressed as the
expectation of a probability measure supported on the set $\{x_{t+\tau} =
[U^\tau X](\omega_t), \; ~\mathrm{for~all}~ \; \omega_t \in B_t\}$---the instantaneous predictive distribution. Although we directly formulated the instantaneous predictive distribution in Eq. (\ref{eq:instant_pred_dist}) using the nonequilibrium measure $\mu_t$, these predictive distributions cannot be directly formulated as an $L^2(\mu_*)$ measure \cite{berr20a}. That said, they provide insight into the intrinsic error of $Z^\tau$, regardless of which measure is used for the nonlinear projection of Eq. (\ref{eq:target_estimator}). As $Z^\tau$ is the expectation of this distribution, the intrinsic error is then seen as the variance of the predictive distribution. Later, we will give explicit constructions of history-dependent generalizations of predictive distributions.



\paragraph*{Empirical Hilbert Spaces}
While $L^2(\mu_*)$ is theoretically convenient, it is not a workable space for
empirical models \cite{gonz21a}. This is because functions in $L^2(\mu_*)$
cannot be distinguished with a finite set of samples, but the latter is what is
empirically available. Data-driven algorithms thus typically employ reproducing
kernel Hilbert spaces (RKHSs), which have well-defined point evaluations. For
more on RKHS methods, as well as empirical sample measures and their
convergence, see Refs. \cite{muan17a,alex20a,berr20a,loom21a}.

\section{Mori-Zwanzig Formalism of Dynamical Processes}
\label{sec:MZ}

Although Eq. (\ref{eq:target_estimator}) defines the optimal instantaneous
model, the optimal model still has an associated intrinsic error---the variance
of the instantaneous predictive distribution. With the same Hilbert space and
projection operator formalism used to define $Z^\tau$, the Mori-Zwanzig
formalism provides the full equations of motion for the observable degrees of freedom by projecting the system dynamics onto those degrees of freedom \cite{chor02a}. The composition of the Mori-Zwanzig equation reveals terms in addition to $Z^\tau$ that lead to the intrinsic error when not accounted for in instantaneous models. Crucially, the additional terms show that partial observation induces a memory dependence in dynamical processes. This then motivates the use of history-dependent models for increased predictive skill over instantaneous models. 

The Mori-Zwanzig setting is identical to that for the partially-observed
dynamical systems considered so far. There is an underlying true system
$(\Omega, \Phi)$ and a noninvertible mapping $X: \Omega \rightarrow
\mathcal{X}$. The variables $x = X(\omega)$ are known as $\omega$'s
\emph{resolved degrees of freedom}. Denoting the remaining \emph{unresolved
degrees of freedom} $\widetilde{x}$, then $\omega = (x, \widetilde{x})$.
The standard formulation of the Mori-Zwanzig equation in statistical mechanics
assumes $(\Omega, \Phi)$ to be Hamiltonian and considers projections of
densities $\rho(\omega)$ and their time evolution by the Liouville operator
\cite{Wild98a}. Importantly, the equation can be derived in our more
general setting of dissipative systems using the Koopman operator
\cite{lin21a,gila21a}.



The goal is to predict the future values of the resolved degrees of freedom
using only information available from them. That is, the task is to express the
evolution of the resolved variables---the dynamics governing the dynamical
process---in terms of the resolved variables as much as possible. We do this by
projecting the Koopman operator's action onto the resolved degrees of freedom.
This is possible since the dynamics of dynamical processes is given in terms of
Koopman operators, as shown above.

Referring Eq.~(\ref{eq:HilbertSpaces})'s Hilbert spaces---i.e., $H = L^2(\mu_*)$---the discrete-time derivation expands the Koopman operator $U^{t+1} : H \rightarrow H$ via the Dyson formula:
\begin{align}
U^{t+1} = \sum\limits_{k=0}^t U^{t-k} PU(QU)^k + (QU)^{t+1}
  ~.
\label{eq:Koopman_expand}
\end{align}
In this, $P$ is an orthogonal projection operator from $H$ to a subspace
$H_\dict \subseteq H_X \subset H$ spanned by basis functions $\dict(x)$ that
depend \emph{only} on the resolved variables $x = X(\omega)$. And, $Q = I - P$
is the orthogonal projection to the unresolved variables.

Recall that $X \in H$ is the observation function that returns data gathered
from measurement recordings of an underlying physical system $\Omega$. Forming
new observable functions in the projected space $H_\dict$ uses observation
measurements in $X$ and functions $\psi(x)$ of them. This is in contrast to
introducing new measurement instruments---instruments that would enlarge the
resolved-variable space $H_X$. In finite subspace projection algorithms, $P$
projects into the subspace $H_\dict$ spanned by the basis of dictionary
functions $\dict$.

In discrete time, the dynamics of the resolved variables are generated via:
\begin{align*}
x_{t+1} & = X(\omega_{t+1}) \\
  & = X_{t+1}(\omega_0) \\
  & = [U^{t+1} X] (\omega_0)
  ~.
\end{align*}
The  discrete-time \emph{Mori-Zwanzig equation} then follows by applying the expansion in Eq. (\ref{eq:Koopman_expand}) to the unit-shift observable $X_{t+1} = U^{t+1}X \in H$.
Skipping algebra and notational
simplifications \cite{gila21a}, this yields:
\begin{align}
x_{t+1} = M_0 (x_t) + \sum\limits_{k=1}^t M_k(x_{t-k}) + \xi_{t+1}(\omega_0)
  ~.
\label{eq:MZ}
\end{align}
The key is that this expression is exact. It gives the true evolution of the
measurement observables, equivalent to the action of the full
infinite-dimensional Koopman operator.

The first term $M_0 (x_t)$ describes Markovian evolution. It gives the best
Markov approximation of $\Phi^1$ under projection $P$. That is, $M_0$ is the
best approximation of the unit-step dynamics by a function of the current
observable only. It is the optimal target function $Z^1$ defined above when the
nonlinear projection given in Eq. (\ref{eq:target_estimator}) is used in Eq.
(\ref{eq:Koopman_expand}). The last term $\xi_{t+1}(\omega_0)$ is the
orthogonal term originating from the initial unresolved components. The second
term captures longer-range statistical dependencies with a discrete convolution
of a memory kernel that depends on the orthogonal terms: $M_k \circ X = P(\xi_k
\circ \Phi) \circ X$. (Statistical mechanics refers to this orthogonal
dependence of  memory as a \emph{fluctuation-dissipation relation}.) All terms
depend on the particular projection operator $P$ used. (For example, the terms
$M_0$ and $\{M_k\}$ may be linear---i.e., matrices---for certain choices of $P$
\cite{lin21b}.)

A comparison is in order: the Mori-Zwanzig perspective of Koopman operator projections
in Eq. (\ref{eq:MZ}) versus the data-driven approaches for finite-dimensional
Galerkin projections of the Koopman operator. 
The latter are presented in Appendix
\ref{app:koopman_appox}; namely, Dynamic Mode Decomposition (DMD) and Extended
Dynamic Mode Decomposition (EDMD). Both DMD and EDMD are instantaneous models
and, as such, only approximate the Markovian term $M_0$. Both do so using
linear finite subspace projections onto $H_\dict \subseteq H_X$. DMD uses the
simple dictionary $\dict = \{f_X\}$ consisting of only the identity function
$f_X$; while EDMD uses arbitrary dictionaries $\dict$ of basis functions. In
contrast, while data-driven approaches to Mori-Zwanzig evolution operators also
use finite subspace projections $H_\dict \subseteq H_X$
\cite{chor02a,chor15a,lin21b}, they do so for the memory kernels as well as for
the Markovian component.

Comparing further, EDMD seeks a Galerkin approximation of the Koopman operator
itself, with a single matrix, while Mori-Zwanzig evolution approximates
projections of the Koopman operator \emph{action} specifically on functions of
the resolved degrees of freedom. Paraphrasing Ref. \cite{lin21b}: ``EDMD seeks
a point $\edmd \dict(x_t)$ in $H_\dict$ that minimizes the error between the
point and $\dict(x_{t+\tau})$, whereas Mori-Zwanzig simply projects
$\dict(x_{t+\tau})$ onto $H_\dict$''.


The crucial insight of the Mori-Zwanzig equation Eq.~(\ref{eq:MZ}) is that
partially observing dynamical systems induces a memory dependence in the
observable degrees of freedom. 
Optimal instantaneous models are thus not fully optimal as data-driven Hilbert space models. 
History-dependent models will reduce intrinsic error and improve predictive skill.
Moreover, the dependence of the memory
kernels on the orthogonal unresolved degrees of freedom indicates that the
memory dependence accounts for the effects of the unresolved variables on the
dynamics of the resolved variables. Recall that much of the effort in
physics-based models comes in explicitly inferring the unobserved degrees of
freedom and their dynamical effects.

\section{Delay-Coordinate Embeddings}
\label{sec:delay_embeddings}

A key step in bridging physics-based and data-driven approaches to prediction
comes through the formulation of memory as reconstruction embeddings \cite{Saue91a}. Their intrinsic geometry illuminates how memory of partial observations implicitly encodes effects of the unobserved degrees of freedom.

Starting with a scalar time series $\{x_t, t \in
\mathbb{N}^+\}$, the task is to reconstruct an effective state space of
\emph{embedding dimension} $m$ in which the effective states evolve as a deterministic
dynamical system. In short, $m$ is set large enough that the orbits in the
reconstructed state space do not intersect. A \emph{derivative-coordinate
embedding} of a measurement observable $x_t$ develops a reconstructed state
space from $\past^m_t = \{\dot{x}_t, \ddot{x}_t, \ldots, d^m x_t / dt^m\}$
\cite{Pack80}. A \emph{delay-coordinate embedding} uses $\past^{m, \delta}_t =
\{x_t, x_{t-\delta}, x_{t-2\delta}, \ldots, x_{t-(m-1)\delta}\}$ with lag
$\delta$ \cite{Take81}. Due to its familiarity we discuss delay-coordinate
embeddings, despite the extra required optimization over lag $\delta$ that may
be required in practice. Unless otherwise stated, we take $\delta = 1$. For
continuous-time systems the lag is given in units of the measurement sample rate $\Delta t$.

The original work on coordinate embeddings established that the geometry of the
asymptotic attractor of $(\Omega, \Phi)$ can be reconstructed, up to
diffeomorphism, from embeddings $\past^{m}$ of partial observations $x =
X(\omega)$ for sufficiently large $m$ \cite{Pack80,Take81}. The intuitive idea
is that the additional values in the embedding $\past^m$ essentially act to
fill in the degrees of freedom of $\omega$ missing from $X$. The reconstructed
orbit $\past_t^m$ of the embedding traces out an attractor that is
geometrically equivalent to that generated by the full system state $\omega_t$.

Geometrically, embeddings encode the unobserved degrees of freedom in the
histories of the observed degrees of freedom. Moreover, the Koopman operator
acting on a delay embedding observables implicitly encodes the unobserved degrees of
freedom in a dynamically useful way \cite{arba17a,gian19a,kamb20a}. In fact,
the Koopman operator acting on delay-coordinate embeddings corresponds to the
Laplace-Beltrami operator describing the attractor geometry \cite{gian19a}.
In the asymptotic limit with evolution on the attractor, this correspondence allows
employing geometric tools, such as heat kernels and diffusion maps, in
data-driven modeling \cite{berr20a}.

\begin{figure}
\begin{center}
\includegraphics[width=0.48 \textwidth]{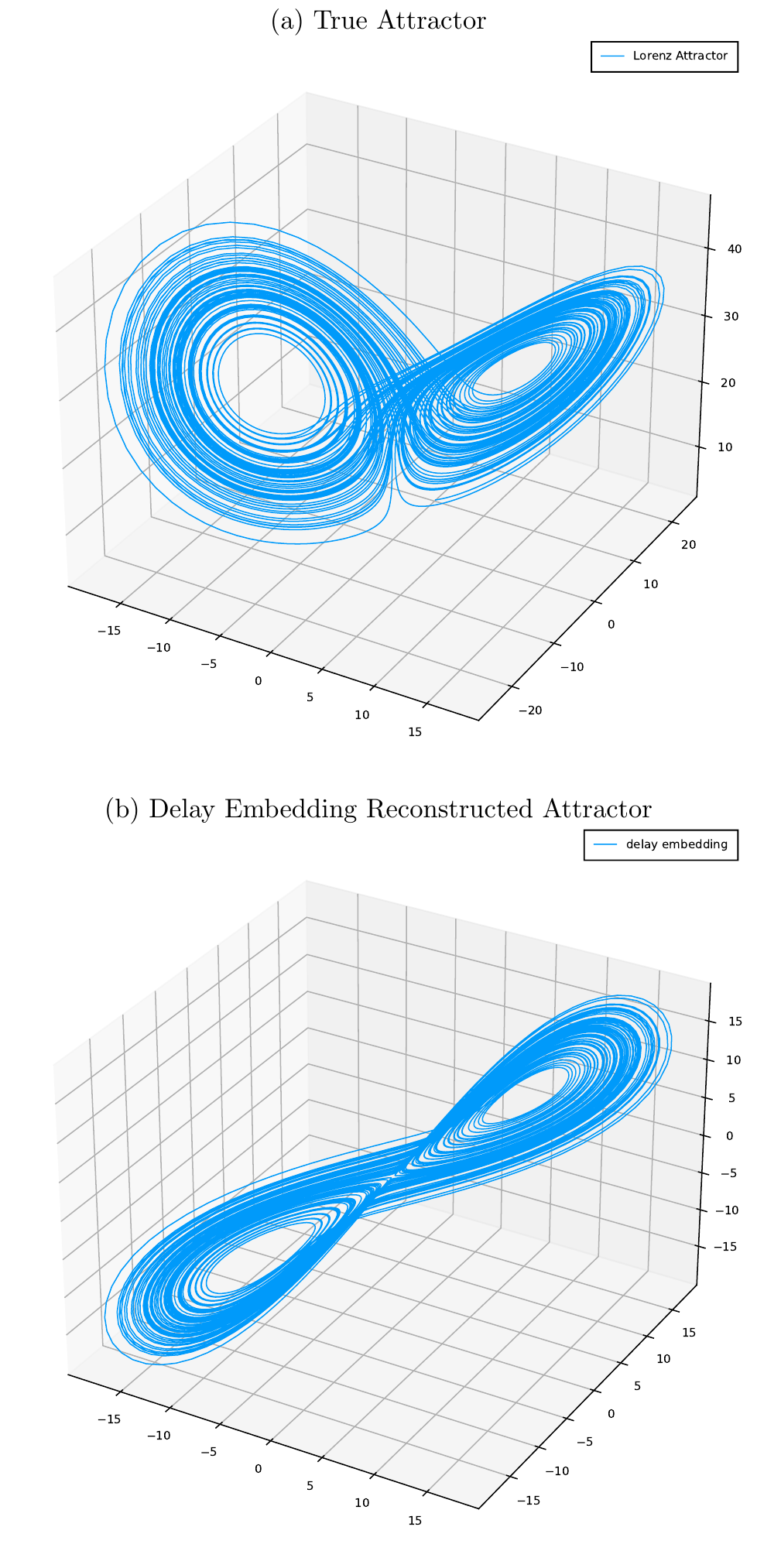}
\end{center}
\caption{(a) Full 3D attractor from numerical solutions of Lorenz equations.
	(b) Delay-embedding reconstruction attractor using $x$ variable only with
	embedding dimension $m = 7$ and lag $\delta = 4$. The first three dimensions $\{x_t, x_{t-4}, x_{t-8}\}$ are shown.
	}
\label{fig:attractors}
\end{figure}

The details of how evolution operators acting on delay-coordinate embeddings dynamically encode the unobserved degrees of freedom will be examined thoroughly below. First though, we show the classical example of how delay embeddings geometrically encode unobserved degrees of freedom with the Lorenz 63 attractor. Later, we will return to this example to demonstrate the dynamical encoding of the unobserved degrees of freedom using analog forecasting.

\paragraph*{Example Reconstruction}
The following gives an empirical demonstration that delay embeddings can geometrically ``fill
in the gaps'' of missing degrees of freedom the three-dimensional Lorenz 63 system:
\begin{align*}
    \dot{x} &= \sigma(y - x)\\
    \dot{y} &= x(\rho -z) - y\\
    \dot{z} &= xy - \beta z
~.
\end{align*}
Figure~\ref{fig:attractors}(a) shows the attractor revealed by their numerical
solution with parameters $\sigma = 10.0$, $\rho=28.0$, and $\beta = 8/3$. For
comparison, Fig.~\ref{fig:attractors}(b) shows the attractor reconstructed
using delay-coordinate embeddings of the $x$ variable alone with $\delta = 4$
and $m = 7$. The delay-reconstructed attractor is a ``squished'' version of the
original, but is (approximately) geometrically equivalent. The simulation and
delay embedding reconstruction were performed using the DynamicalSystems.jl
package in Julia \cite{Dats18a}.

\section{History-Dependent Models}
\label{sec:history-dep_models}

Many history-dependent model classes, such as recurrent neural networks
\cite{chat20a}
, are
more readily understood as mappings $\overleftarrow{\mathcal{T}}^\tau$ from
pasts (delay embeddings) to future observations, rather than as fitting the
paradigms of Markov and memory kernels from the Mori-Zwanzig equation. Generalizing the instantaneous case
given in Eq. (\ref{eq:iddmodel}), the mean-squared error of history-dependent
Hilbert space models is:
\begin{align}
\|\overleftarrow{\mathcal{T}}^\tau \circ \Past^k - U^\tau X\|^2_{L^2(\mu_*)}
  ~.
  \label{eq:hddmodel}
\end{align}
As in the instantaneous case, we identify the unique minimizer of Eq.
(\ref{eq:hddmodel}) as the optimal history-dependent Hilbert space model. This optimum is achieved by formally connecting the Mori-Zwanzig formalism with
delay-coordinate embeddings using Wiener projections of the Koopman operator.

Before detailing Wiener projections, it is helpful to first review two standard
orthogonal projections. Both use the $L^2(\mu)$ inner product. For now, we
follow Ref. \cite{lin21a} and use the invariant measure $\mu_*$:
\begin{align*}
    \langle f, g\rangle_{L^2(\mu_*)} = \int_\Omega f \; g \; d\mu_*
    ~.
\end{align*}

Equation (\ref{eq:target_estimator}) defined the optimal instantaneous model
$Z^\tau$ as the conditional expectation function that minimizes the
$L^2(\mu_*)$ norm between $g \circ X$ and $U^\tau X$. This is known as the
nonlinear or \emph{infinite-rank} projection, used by Ref. \cite{zwan01a}, of
$U^\tau X$ from $H$ into $H_X$.

In contrast, the linear projection used by Ref. \cite{mori65a}, also known as a
\emph{finite-rank} or finite-subspace projection, is defined in terms of an
orthogonal set $\dict$ of size $N$ on the space $V$ of functions of the
observed variables. That is:
\begin{align}
P f = \sum_{i=1}^N \langle f, \phi_i \rangle_{L^2(\mu_*)} \; \psi_i
    ~,
\label{eq:lin_proj}
\end{align}
where $\{\phi_i = \psi_i \circ X\}_{i=1}^N$.

In the infinite-rank limit $\dict \rightarrow V$, this linear projection
converges to the nonlinear conditional expectation projection. As with EDMD,
the challenge for data-driven methods that employ finite-subspace projections
\cite{zhao16a,alex20a,gila21a,lin21b} is to find an effective finite basis
$\dict$.

\subsection{Equilibrium Wiener Projections}

Using these instantaneous projections, the standard Mori-Zwanzig formalism
embodies history dependence of the observed variables in the collection of memory kernels.
In contrast, \emph{Wiener projections} incorporate history dependence directly
into the projection operators via delay-coordinate embeddings. Specifically,
the linear Wiener projections replace the single $L^2(\mu_*)$ inner product
with:
\begin{align}
\langle f, g \rangle_W := \lim\limits_{k\rightarrow \infty} \frac{1}{k} \sum\limits_{\tau=1}^k \int_\Omega [U^{-\tau}f] [U^{-\tau}g] d\mu_*
  ~,
\label{eq:Wiener_proj}
\end{align}
in the linear projection in Eq.~(\ref{eq:lin_proj}). That is, $L^2(\mu_*)$
inner products of the reverse-time-shifted observables are taken at all times
into the infinite past.


Applying a discrete-time Wiener projection $P_W$ to the discrete-time
Koopman operator, as first introduced in Ref. \cite{lin21a}, results in:
\begin{align}
U^{t+1} = U^t P_W U + (Q_W U)^{t+1}
  ~.
\label{eq:Koop_delay_expand}
\end{align}
As expected, there is no longer a temporal convolution over memory kernels;
only a Markov term and an orthogonal term. In the setting of quantum
statistical mechanics, Ref. \cite{koid02a} gives a similar expression using
time-dependent projection operators.

Furthermore, Ref. \cite{gila21a} argues that, if the conditions of the delay
embedding theorem \cite{Take81} are met, the orthogonal term vanishes. Thus,
\emph{in the ideal case}, Mori-Zwanzig evolution with delay-coordinate
embeddings reduces to only a single Markov term that corresponds to the
nonlinear Wiener projection of $U^{t+1} X$:
\begin{align}
x_{t+1} & = \mathbb{E}[U^{t+1} X | \Past] \nonumber \\
  & = \markovOP(\past_t)
  ~,
\label{eq:markovOP}
\end{align}
for embedding dimension $m$ sufficiently large to satisfy the delay-embedding theorem.

In the nonideal case, particularly with finite $k$, the orthogonal term does not vanish and there may be memory effects at Markov order larger than that spanned by finite pasts $\past_t^k$. Therefore, as with standard
(instantaneous) Mori-Zwanzig Markov approximations, the stochastic evolution
of finite pasts is modeled by the finite Markov operator $\markovOP^k(\past^k)$
plus an effective ``noise'' term:
\begin{align}
\Pr(X_{t+1} | \Past^k=\past^k) = \markovOP^k(\past^k) + \text{noise}
  ~.
\label{eq:finite_markovOP}
\end{align}
A finite model of this form is found in the HAVOK method
\cite{brun17a,kamb20a}, based on Hankel DMD \cite{arba17a}. This finds the
best-fit linear approximation for $\markovOP^k$ with the leading components of
the singular value decomposition of the Hankel matrix, whose columns are
time-ordered delay embeddings. The last few components are then fit to the
noise.

We emphasize that the Wiener projection approach to Mori-Zwanzig evolution is
useful as it provides a direct connection to delay-coordinate embeddings and their intrinsic geometry.
Theoretically, though, it merely rearranges memory dependence in the observable
degrees of freedom. Delineating the practical advantages or disadvantages of
Wiener projections over the standard Mori-Zwanzig formalism requires further
investigation. Note, though, that the algorithms given by Ref. \cite{lin21b}
for reconstructing the Markov and memory kernels of the latter employ two-time
correlation functions of the observed variables. These are closely related
to the instantaneous predictive distributions described above.

\subsection{Nonequilibrium Wiener Projections}

For another perspective on how the orthogonal term in Eq.
(\ref{eq:Koop_delay_expand}) may vanish, it is instructive to formulate Wiener
projections using nonequilibrium time-dependent measures, rather than the
equilibrium invariant measure. In statistical mechanics, in fact, the invariant
measure is taken to be exactly that of the equilibrium distribution, with the
$L^2$ inner products being equilibrium correlations and the Mori-Zwanzig
equation's validity holding only near equilibrium \cite{te19a}.

First, we introduce the history-dependent generalization of the time-dependent
Maximum Entropy measures introduced in Section~\ref{sec:physPO}. The
history-dependent generalization of the Maximum Entropy Principle is known as
Maximum Caliber \cite{Jayn85a,Gran08a}. In short, given a time series of
constraints up to the present moment, Maximum Caliber constructs the least
biased distribution at the current time by maximizing the entropy while
accommodating all time-evolving constraints. If the constraints are given in
the form of expectation values, as is typical in statistical mechanics, this
results in generally intractable spacetime path integrals.

As in the instantaneous case though, constraints for partially-observed systems are simply support sets of possible $\omega$ consistent with observations $x=X(\omega)$.
Now, however, there are multiple time-evolving
observations $\{x_t, x_{t-1}, \ldots, x_{t-k}\}$ in the form of delay
embeddings to constrain the support sets over $\Omega$.

Equation~(\ref{eq:instantaneous_B}) defined the instantaneous set $B_t \in
\Sigma_\Omega$ as the set $X^{-1}(x_t)$ of all $\omega_t$ consistent with the
observation $x_t$ such that $X(\omega_t) = x_t$. Rather than a single
instantaneous observation, consider now two sequential observations $x_{t-1}$
and $x_t$. Define $\overleftarrow{B}_t^2 \in \Sigma_\Omega$ as the set of
$\omega_t$ consistent with both observations such that $X(\omega_t) = x_t$ \emph{and} $[U^{-1}X](\omega_t) = X(\Phi^{-1}(\omega_t)) = x_{t-1}$.
Note that $\overleftarrow{B}_t^2 \subseteq B_t$. If the two sets are not equal, we say that $\overleftarrow{B}_t^2 = B_t \cap X^{-1}(x_{t-1})$ \emph{refines} $B_t$.

\begin{figure*}
\begin{center}
\includegraphics[width=0.9 \textwidth]{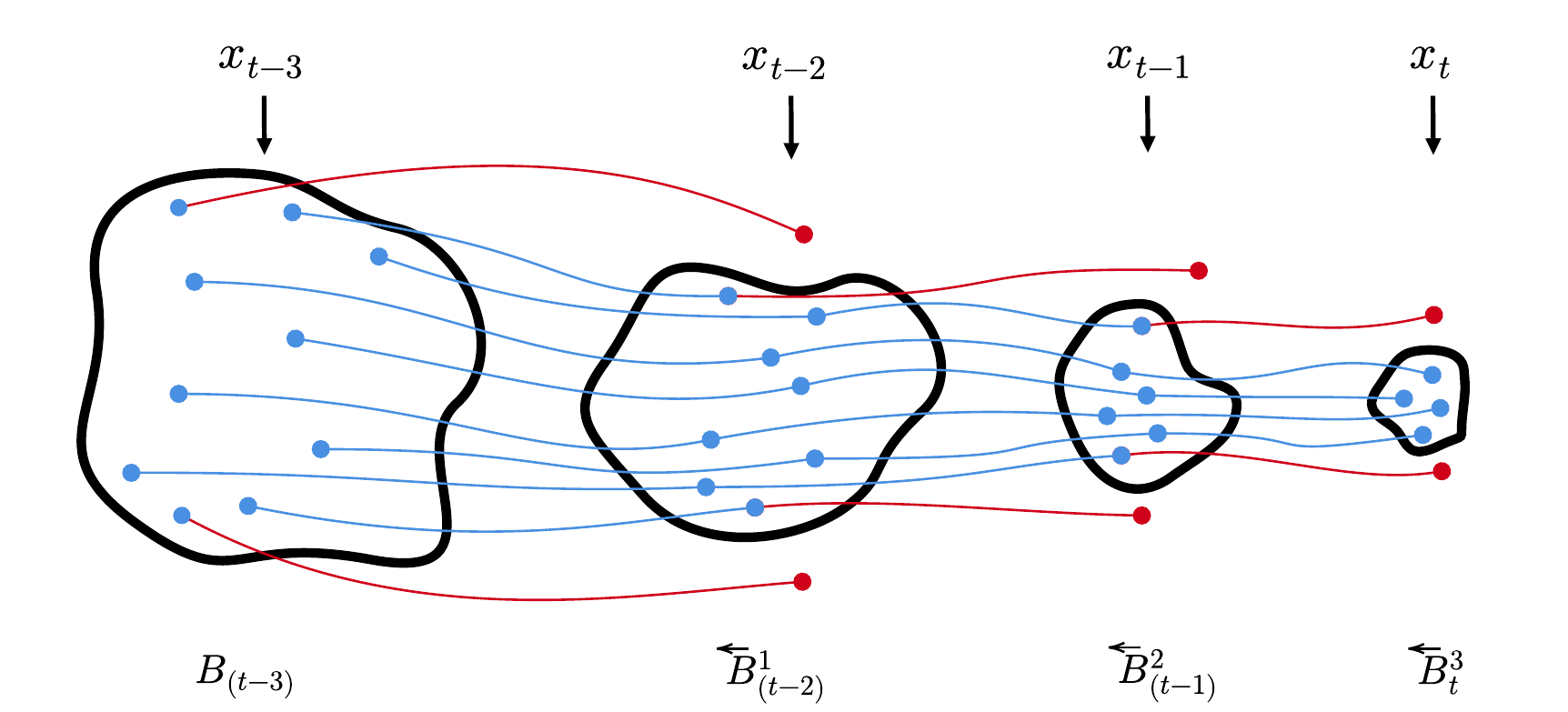}
\end{center}
\caption{Sequential support sets of Maximum Caliber measures for past lengths
	$k = \{0,1,2,3\}$: The earliest observation $x_{t-3}$ alone produces the
	$k=0$ support set $B_{(t-3)} = X^{-1}(x_{t-3})$. The next observation
	$x_{t-2}$, together with $x_{t-3}$ yields
	$\protect\overleftarrow{B}_{(t-2)}^1 = B_{(t-3)} \cap X^{-1}(x_{t-2})$.
	Similarly, $x_{t-1}$ gives $\protect\overleftarrow{B}_{(t-1)}^2 =
	\protect\overleftarrow{B}_{(t-2)}^1 \cap X^{-1}(x_{t-1})$ and, finally,
	$x_t$ gives $\protect\overleftarrow{B}_{t}^3 =
	\protect\overleftarrow{B}_{(t-1)}^2 \cap X^{-1}(x_{t-1})$. Unit-length
	orbits of $\omega_\tau \in \Omega$ consistent with the subsequent
	observation $x_{\tau+1}$ are shown in blue, while those inconsistent with
	the subsequent observation are red. The depiction shows strict refinement
	at each time, so that $\protect\overleftarrow{B}_t^3 \subset
	\protect\overleftarrow{B}_{(t-1)}^2 \subset
	\protect\overleftarrow{B}_{(t-2)}^1 \subset B_{(t-3)}$.
	}
\label{fig:maxcal-supps}
\end{figure*}

For a depth-$k$ past $\past_t^k = \{x_t, x_{t-1}, \ldots,
x_{t-k}\}$---a $k+1$-dimensional delay embedding---we define the
set $\overleftarrow{B}_t^k$ as:
\begin{align}
\overleftarrow{B}_t^k := \{\omega_t \; | \; & X(\omega_t) = x_t,
    [U^{-1}X](\omega_t) = x_{t-1}, \ldots, \nonumber \\
    & \qquad [U^{-k}X](\omega_t) = x_{t-k}\}
    ~.
    \label{eq:hist_depend_B}
\end{align}
Given the set $\overleftarrow{B}_t^k$, the Maximum Caliber distribution is uniform over $\overleftarrow{B}_t^k$ and zero elsewhere, as with the instantaneous Maximum Entropy case. For finite $k$, $\overleftarrow{B}_t^k$ is $\nu$-measurable and the Maximum Caliber measure $\overleftarrow{\mu}_t^k$ is defined through the density $\overleftarrow{\rho}_t^k$ that is constant over $\overleftarrow{B}_t^k$ and zero elsewhere. If $\lim_{k\rightarrow \infty}\overleftarrow{B}_t^k$ is a discrete set, $\overleftarrow{\mu}_t^\infty$ is given as a sum of equally-weighted delta distributions.


With the Maximum Caliber measures in hand, we can define nonequilibrium
Wiener projections using the nonequilibrium inner products:
\begin{align}
    \langle f_t, g_t \rangle_{\past_t^k} := \frac{1}{k} \sum\limits_{\tau=1}^k \int_\Omega [U^{-\tau}f_t] [U^{-\tau}g_t] d\overleftarrow{\mu}_t^k
    ~,
    \label{eq:noneq_innerprod}
\end{align}
with $f_t = f(\omega_t)$ and $g_t = g(\omega_t)$ to emphasize that the
integrals are all carried out over values of $\omega_t$ for all terms in the
sum. Note that this inner product is identical to its equilibrium counterpart
in Eq. (\ref{eq:Wiener_proj}) except for the change of measure. The conditional
expectation in Eq. (\ref{eq:markovOP}) can be similarly defined using
nonequilibrium Wiener projections defined by Eq. (\ref{eq:noneq_innerprod})'s
inner product.

Unlike Eq. (\ref{eq:Wiener_proj})'s equilibrium case, Eq.
(\ref{eq:noneq_innerprod})'s inner product is contingent on the observation
$\past_t^k$ that then defines the measure $\overleftarrow{\mu}_t^k$. In
particular, we can analyze the sequential refinement behavior of $\overleftarrow{\mu}_t^k$ and
their support sets $\overleftarrow{B}_t^k$ with increasing depth $k$ of
the observed past $\past_t^k$. This is shown in Fig.~\ref{fig:maxcal-supps}. Moreover, we can directly construct the
predictive distributions whose expectation gives Eq. (\ref{eq:markovOP})'s conditional expectation using the (nonlinear) nonequilibrium Wiener projections.

\begin{figure*}
\begin{center}
\includegraphics[width=0.95 \textwidth]{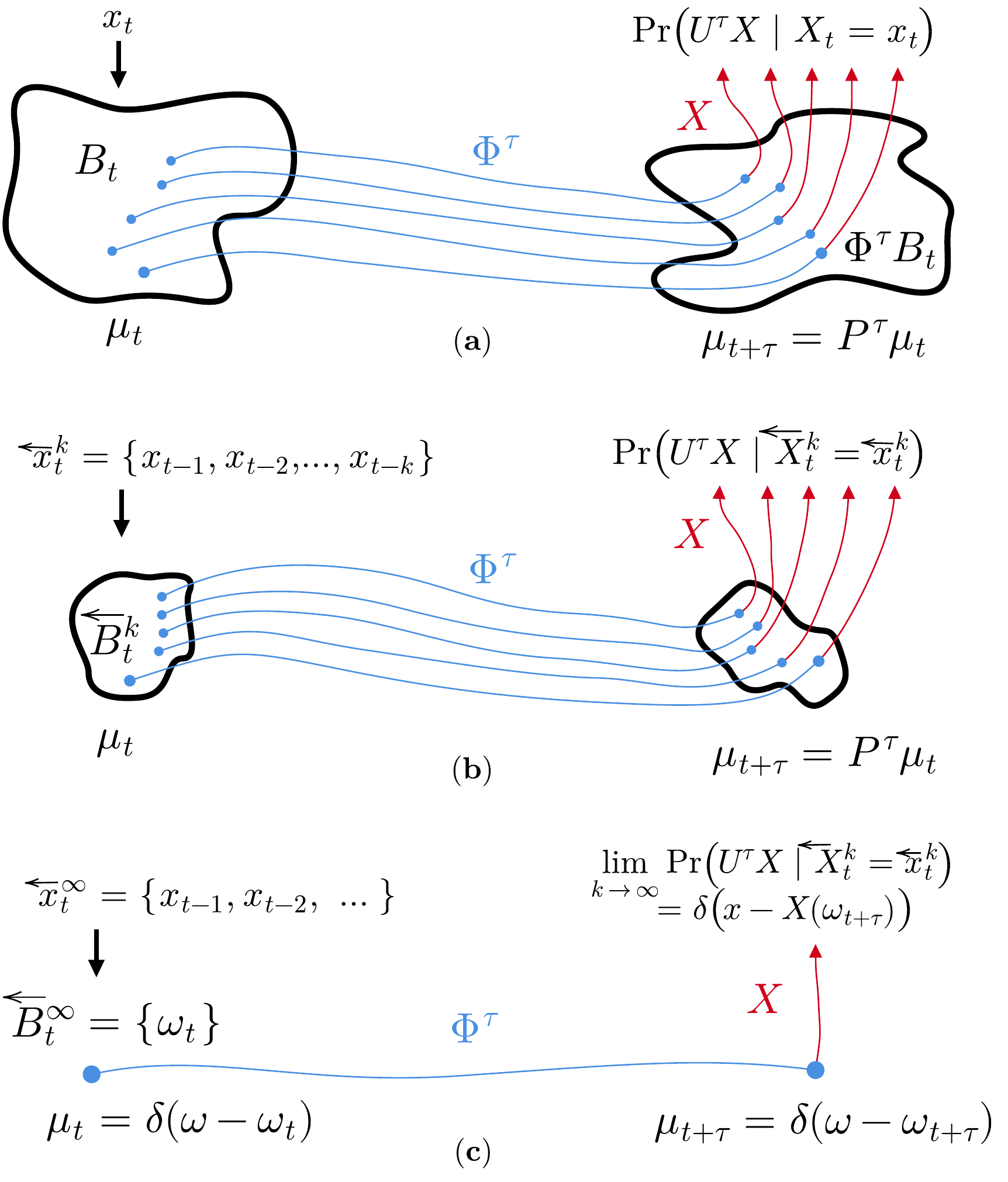}
\end{center}
\caption{Operator-theoretic construction of predictive distributions: (a)
	Instantaneous case, with depth-$k=0$ past consisting of a single
	observation $x_t$. (b) Intermediate case, with depth-$k$ past
	$\protect\past_t^k$. (c) Limiting case, with $\lim k \rightarrow \infty$
	depth past $\protect\past_t^\infty$. In all cases, the initial measure
	$\mu_t$ is uniform on the set of all $\omega_t$
	consistent with observation $\protect\past_t^k$. The initial measure is
	then propagated forward in time with the Perron-Frobenius operator
	$P^\tau$ and, finally, the predictive distribution is given according to the
	pushforward of $\mu_{t+\tau} = P^\tau \mu_t$ along the observable mapping
	$X$. (c) depicts the case when there is one and only one initial Platonic
	state $\omega_t$ consistent with the $\infty$-length observation
	$\protect\past_t^\infty$. Then, the prediction converges to the true
	evolution $x_{t+\tau} = X\bigl(\Phi^\tau(\omega_t)\bigr) = [U^\tau
	X](\omega_t)$.
}

\label{fig:pred_dist}
\end{figure*}

\subsection{Predictive Distributions}

In the instantaneous case, recall that $B_t = X^{-1}(x_t)$ is the set of all
$\omega_t$ consistent with observation $x_t$ such that $X(x_t) = \omega_t$.
The set of possible observables $x_{t+\tau}$ that may be seen at a later time
$\tau$ are given by the Koopman operator as $\{x_{t+\tau} =
[U^{\tau}X](\omega_t) ~\mathrm{for~all}~ \omega_t \in B_t\}$. This set is
the support of the instantaneous predictive distribution $\mu_{t+\tau}^X$---the
pushforward of $P^{\tau}\mu_t$ along $X$.

For two sequential observations, we may expect that there are some, if not
many, $\omega_t$ in $B_t$ that are not in $\overleftarrow{B}_t^2$. That is,
there may be $\omega_t$ such that $X(\omega_t) = x_t$ but
$X\bigl(\Phi^{-1}(\omega_t)\bigr) \neq x_{t-1}$. As more observations are
recorded, there may be increasingly fewer $\omega_t$ whose reverse orbit is
consistent with the observed values $\past_t^k$. Therefore,
$\overleftarrow{B}_t^k \subseteq B_t$ and, in some cases,
$\overleftarrow{B}_t^k$ is a proper subset of $B_t$ with
$\nu(\overleftarrow{B}_t^k) < \nu(B_t)$. That is, the state space volume
$\nu(\overleftarrow{B}_t^k)$ is monotonically nonincreasing as $k$ grows and it
may decrease for increasing $k$. Figure~\ref{fig:maxcal-supps} illustrates the
dynamical refinement of Maximum Caliber support sets.

We are now ready to examine the consequences of refinement on history-dependent
predictive distributions. The latter are constructed as for instantaneous
predictive distributions, using uniform initial measures over
$\overleftarrow{B}_t^k$ rather than $B_t$. The \emph{predictive distribution}
$\Pr(U^{t+\tau}X | \Past_t^k = \past_t^k)$ is supported on the set
$\{x_{t+\tau} = [U^\tau X](\omega_t) ~\mathrm{for~all}~ \; \omega_t \in
\overleftarrow{B}_t^k\}$ and is distributed according to the pushforward of
$P^\tau \overleftarrow{\mu}_t^k$ along $X$, as shown in
Fig.~\ref{fig:pred_dist}~(b).

If $\nu(\overleftarrow{B}_t^k) < \nu(B_t)$, then the initial measure
$\overleftarrow{\mu}_t^k$ is more constrained than $\mu_t$. And so, $P^\tau
\overleftarrow{\mu}_t^k$ is also be more constrained than $P^\tau \mu_t$. In
this case, the predictive distribution $\Pr(U^{t+\tau}X | \Past_t^k =
\past_t^k)$ has lower entropy than the instantaneous $\Pr(U^{t+\tau}X | X_t =
x_t)$. Here, ``entropy'' refers the size of a distribution's support.  The
volume of $\{x_{t+\tau} = [U^\tau X](\omega_t) ~\mathrm{for~all}~ \omega_t \in
\overleftarrow{B}_t^k\}$ is no larger than that of its instantaneous
counterpart $\{x_{t+\tau} = [U^{\tau}X](\omega_t) ~\mathrm{for~all}~ \omega_t
\in B_t\}$ since $\overleftarrow{B}_t^k \subseteq B_t$ and $U^\tau X$ is
measurable. Note that for distributions with a density $\rho$, the size of the
effective support set---the \emph{typical set}---is given by $2^{h(\rho)}$,
where $h(\rho)$ is the differential entropy of $\rho$ \cite{Cove91a}.

Taking $\tau=1$, consider the optimal history-dependent target function
$\overleftarrow{Z}^k = \mathbb{E}[U^{t+1}X | \Past_t^k = \past_t^k]$ and the
instantaneous optimal $Z = \mathbb{E}[U^{t+1}X | X_t = x_t]$. From the
arguments above, $\nu(\overleftarrow{B}_t^k) < \nu(B_t)$ implies that
$\overleftarrow{Z}^k$ is a more accurate estimator of $x_{t+1} =
[U^{t+1}X](\omega_t)$ than the instantaneous optimal $Z$. This follows since
there is less variance in $\Pr(U^{t+1}X | \Past_t^k = \past_t^k)$---it is more
tightly concentrated about its mean than $\Pr(U^{t+1}X | X_t = x_t)$. Such a
conclusion is in line with the intuition that the intrinsic error of
instantaneous models derives from the unobserved degrees of freedom and that
including past observations in the form of delay-coordinate embeddings accounts
for the missing degrees of freedom.

It is natural to ask what happens in the $k \rightarrow \infty$ limit of
infinitely-many past observations. Reference \cite{gila21a} concludes that, for
sufficiently large $k$, the estimator $\mathbb{E}[U^{t+1}X | \Past^t] =
X_{t+1}$ is the identity map that yields the true evolution of $x_{t+1}$. In
the nonequilibrium case, this implies that, as $k \rightarrow \infty$, the size
$\nu(\overleftarrow{B}_t^\infty)$ vanishes, with only a single $\omega_t$
consistent with the infinite set of observations in $\past_t^\infty$. As a
consequence, there is a unique value $x_{t+1} = [U^1 X](\omega_t)$ for $\omega_t \in \overleftarrow{B}^\infty_t=\{\omega_t\}$.
Similarly, $\overleftarrow{\mu}_t^\infty$ is a $\delta$-distribution at
$\omega_t$ in this case, with $\overleftarrow{\mu}_{t+1}^\infty =
P^{1}\overleftarrow{\mu}_t^\infty$ also a $\delta$-distribution at
$\omega_{t+1} = \Phi^{1}(\omega_t)$. The conditional distribution $\Pr(U^{t+1}X
| \Past_t^\infty \past_t^\infty)$ is then a $\delta$-distribution with support
on the single $x_{t+1} = X\bigl(\Phi^1(\omega_t)\bigr)$. See
Fig.~\ref{fig:pred_dist}(c).


This clearly shows how coordinate embeddings recover the unobserved degrees
of freedom and effectively act as an equivalent to the full underlying Platonic
state $\omega$. In the ideal case, as the length of an embedding increases to
infinity, the corresponding size of the set $\overleftarrow{B}_t^k$ of possible
initial conditions goes to zero, as does the variance of the resulting predictive distribution. Thus, there is an a.e. one-to-one correspondence between infinite-length delay embeddings and Platonic system states $\omega$. 
The associated predictive distribution converges, in a thermodynamic limit, to the true value of the next observable, given by the Wiener projection of $U^{t+1}X$.

Consider, for example, the simple harmonic oscillator. The underlying state
$\omega$ is two-dimensional: $\omega = (r, p)$. Assume access only to position:
$X(\omega) = r$. The evolution of position is described by sine waves $r(t) =
\sin(t)$. At a given time instant, we cannot determine the momentum from the
instantaneous position alone and, therefore, cannot determine the full
underlying state $\omega$. However, there are only two momenta associated with each
instantaneous position---call them \emph{positive} and \emph{negative}.
Following Fig.~\ref{fig:pred_dist}'s procedure for constructing predictive
distributions, there are two corresponding system states and so two
corresponding future position values at time $t+dt$. Call these \emph{up} (for
positive momentum) and \emph{down} (for negative momentum). If we include a
single infinitesimal past value from time $t-dt$, this is a.e. sufficient to
distinguish if the momentum term is positive or negative at that time. Thus,
except for the turning points (that are measure zero), including that single
past value produces a $\delta$-distribution for the inferred $\mu_0$ and, thus,
gives the exact future prediction using Eq.~(\ref{eq:markovOP}).

This convergence, however, is not guaranteed. The size of
$\overleftarrow{B}_t^k$ \emph{may} decrease with increasing $k$, but it does
not necessarily always do so. Interestingly, while chaotic instabilities make
future predictions challenging, they actually aid in this convergence.
Trajectory divergence in forward time \cite{meiss07a} means convergence in
reverse time. Generating partitions $X_{\mathbb{G}}$ of symbolic processes,
detailed in Appendix~\ref{app:gen_part}, are a rigorous case where this is
known to hold \cite{Lind95a}. In that setting, $\overleftarrow{B}_t^k$ is the
element of the dynamical refinement of the generating partition corresponding
to the observed symbol sequence $\past_t^k$. In the limit of infinitely-many
observations the size $\nu(\overleftarrow{B}_t^k)$ of the refined partition
elements vanishes and almost-every infinite-length symbol sequence corresponds
to a unique system state $\omega \in [0,1]$. Generating partitions on chaotic maps of the unit interval are a rigorous case where the a.e. convergence of $\overleftarrow{B}_t^\infty$ to a single $\{\omega_t\}$ is achieved. 

\section{Supporting Examples}
\label{sec:examples}
We now provide examples to demonstrate the ability of delay-coordinate embeddings and Wiener projections of Koopman operators to dynamically encode unobserved degrees of freedom in practice. The data-driven models for these demonstrations employ analog forecasting, Eq.~(\ref{eq:analog_forecast}), due to its simplicity and flexibility. Various classes of analog forecasting target functions are formed based on what inputs are given, with appropriate distances computed to find the analog. Fully-observed instantaneous, partially-observed instantaneous, and partially-observed history-dependent target functions will all be considered. 

\begin{figure}
\begin{center}
\includegraphics[width=0.48 \textwidth]{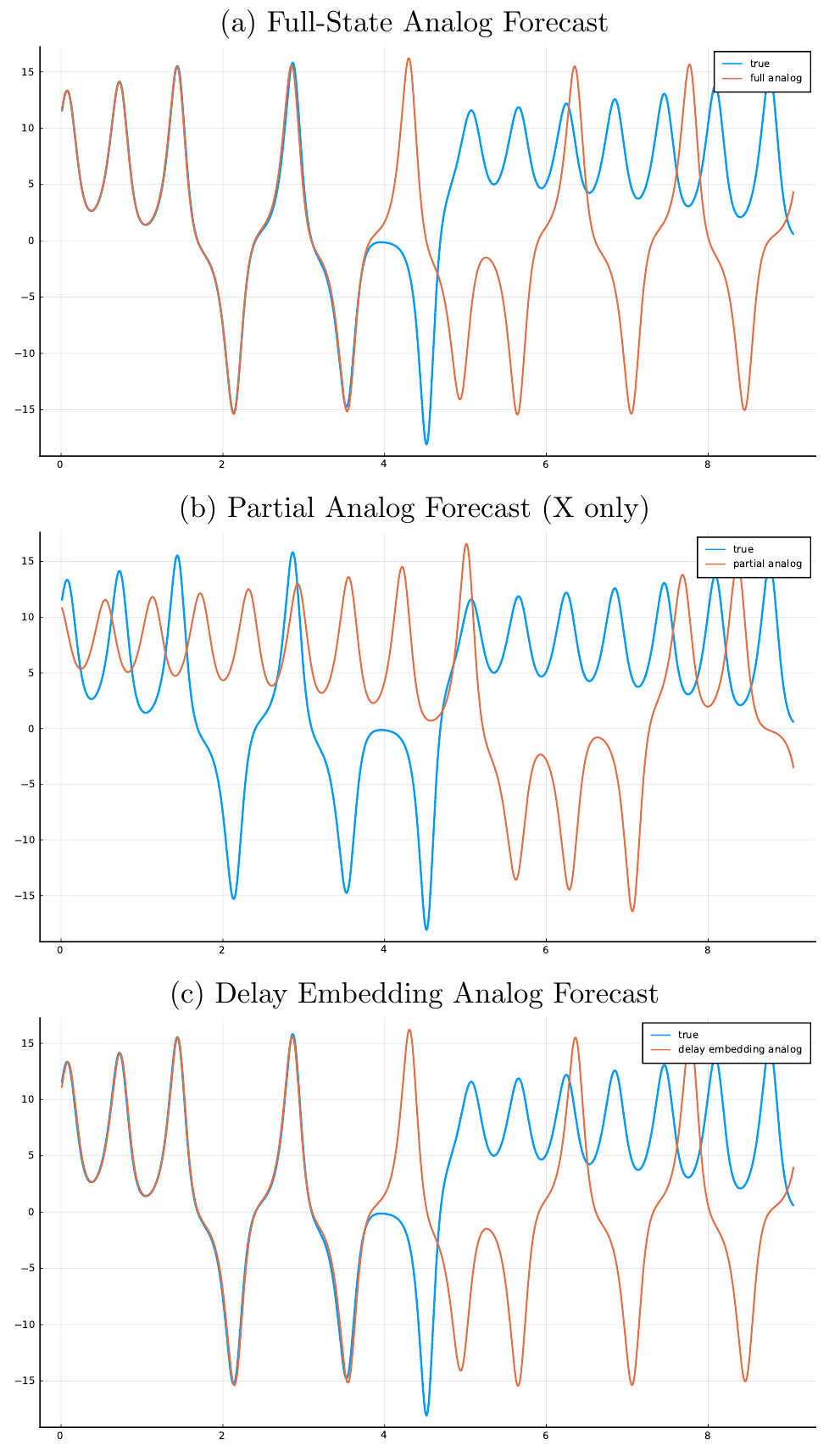}
\end{center}
\caption{Analog forecasting the Lorenz 63 system using different prediction
	variables. (a) All three Lorenz variables $(x,y,z)$ are used to compute
	the analog, with only the trajectory of the $x$ variable shown. (b) The
	$x$ variable alone is used to compute the analog. (c) Delay embeddings
	of the $x$ variables alone, using the same embedding parameters $\delta =
	4$ and $m=7$ used in Fig.~\ref{fig:attractors}(b). The forecast using
	delay-embeddings in (c) is essentially identical to the forecast using the
	full system state in (a).
	}
\label{fig:lorenz-analog}
\end{figure}

\subsection{Lorenz 63}
First, Fig.~\ref{fig:lorenz-analog} provides a dynamical complement to Fig.~\ref{fig:attractors}'s geometric demonstration of encoding unobserved degrees of freedom in the Lorenz 63 system. That is, analog forecast target functions for both cases shown to be geometrically equivalent in Fig.~\ref{fig:attractors} have the same predictive skill.

As a baseline, consider the fully-observed case with $X(\omega_t) = \omega_t = (x_t,y_t,z_t)$. Figure~\ref{fig:lorenz-analog}(a) shows an instantaneous analog forecast that employs the full state variable as $\mathcal{T}(x_t, y_t, z_t)$. This is plotted alongside a numerical integration of the equations of motion in units of Lyapunov time. 

For comparison, Fig.~\ref{fig:lorenz-analog}(b) shows an instantaneous analog forecast via
$\mathcal{T}(x_t)$ using only the first coordinate $x$. While the analog
forecast using the full state variable $(x,y,z)$ tracks the numerical
integration for almost four Lyapunov times, the analog forecast using only the
instantaneous $x$ variable diverges immediately. Despite this quantitative
divergence, the analog forecast using only $x$ produces a qualitatively
consistent forecast, capturing behavior expected of the Lorenz system with
oscillations within, and jumps between, the two attractor lobes.

Fig.~\ref{fig:lorenz-analog}(c) shows an analog forecast using
delay-coordinate embeddings $\past^m_t$ of the $x$ variable with
$\mathcal{T}(x_t, x_{t-\delta}, \ldots, x_{t-(m-1)\delta})$. This is, in fact, the same delay
embeddings used above in Fig.~\ref{fig:attractors}(b) with $\delta = 4$ and $m
= 7$. As with the forecast shown in Fig.~\ref{fig:lorenz-analog}(b), the
forecast in (c) only has access to information in the $x$ variable. However, as
can be seen, including information from past observations of $x$, in the form
of delay embeddings, results in a model with essentially the same predictive
skill as the fully-observed case shown in (a), with divergence again occurring
at about four Lyapunov times.

This shows that, at least in simple low-dimensional cases, delay embeddings
fill in the gaps and so become effective proxies for the missing degrees of freedom
in implicit predictive models. As far as we are aware, the equivalence of predictive skill between an instantaneous full-state model and a history-dependent partially-observed model has not been previously demonstrated.

\subsection{Lorenz 96}
The Lorenz 63 model is useful to connect the geometric encoding of the unobserved degrees of freedom by delay embeddings with the dynamical encoding of the unobserved degrees of freedom by history-dependent models. However, it is a low-dimensional system, so that even when just a single degree of freedom is accessible, there are only two that are inaccessible. 

In this example we demonstrate the effects in increasing-length input pasts with a higher-dimensional system using a 50-dimensional Lorenz 96 model. Each degree of freedom $x^i$ in the model evolves according to local interactions as
\begin{align*}
    \frac{dx^i}{dt} = (x^{i+1}-x^{i-2})x^{i-1} - x^i + F
    ~,
\end{align*}
with periodic boundary conditions. We set $F=4.6$ and numerical integration is performed with a time step $\Delta t = 0.01$. Again we use analog forecasting as the data-driven model, which has access to only a single degree of freedom $x_t=X(\omega_t) = x^1_t$.

\begin{figure*}
\begin{center}
\includegraphics[width=1.0 \textwidth]{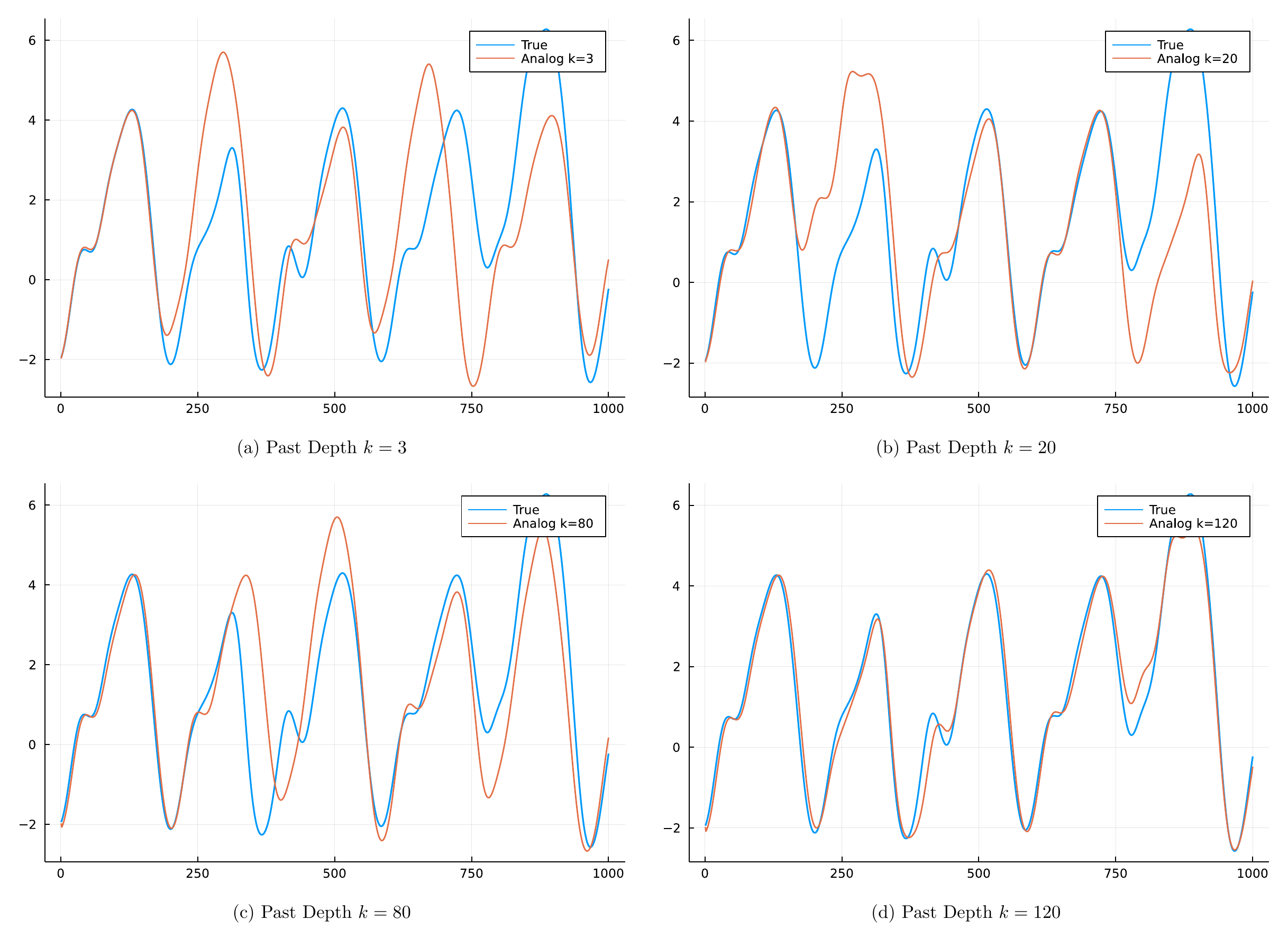}
\end{center}
\caption{Analog forecasting the Lorenz 96 system using input pasts of increasing depth. All predictions are made iteratively with analog forecast target functions of the form $\mathcal{T}(x_t, x_{t-1}, \ldots, x_{t-k})$. Model predictions as a function of integration time-steps are given for four values of past depth $k$.  
	}
\label{fig:lorenz-96}
\end{figure*}

Figure~\ref{fig:lorenz-96} shows four analog forecast predictions of $x^1_t$ in the Lorenz 96 system using history-dependent target functions of the form $\mathcal{T}(x_t, x_{t-1}, \ldots, x_{t-k})$. Each predicion is made with a different value of past depth $k$. As delay-coordinate embeddings, the embedding dimension is $k$ and we use a unit lag $\delta = 1$. Like those in Fig.~\ref{fig:lorenz-analog}, predictions are made iteratively, with the target functions outputting a single prediction at the next time step. 

The lowest-memory prediction is made with $k=3$ and is shown in Fig.~\ref{fig:lorenz-96}(a). The forecast follows the numerical integration for over $200$ integration time steps before diverging. The mean-squared error of the prediction over the $1000$ time-step window is $3.35$. Predictions made with $k=20$ are shown in Fig.~\ref{fig:lorenz-96}(b). While it diverges from the numerical integration sooner than the $k=3$ model, the quasi-periodicity of the Lorenz 96 system allows the forecast to closely track the numerical integration again at later times. The mean-squared error for the $k=20$ model is thus slightly lower, at $3.18$. Increasing the past depth to $k=80$, shown in Fig.~\ref{fig:lorenz-96}(c), provides a more noticeable improvement in predictive skill. This is apparent visually, and is reflected in the mean-squared error value of $1.72$. Finally, increasing to $k=120$, roughly the ideal $2N+1$ embedding dimension, provides a dramatic improvement. The data-driven model closely follows the numerical integration for most of the $1000$ time-step prediction window, giving a mean-squared error of $0.15$. 

Similar results are shown in Fig.~$4$ of Ref.~\cite{gila21a} for a $5$-dimensional Lorenz 96 model with $F=8.0$. The authors use the more sophisticated kernel analog forecasting algorithm, first introduced in Ref.~\cite{zhao16a}.

We emphasize again that convergence of history-dependent data-driven models to the true dynamics is not guaranteed. The results shown in these experiments should not be expected as generic behavior for data-driven models. However, they serve as clear demonstrations of the ability for data-driven models to implicitly encode the effects of unobserved degrees of freedom. This ability provides a physical basis for the efficacy of data-driven predictive models, and provides a bridge connecting them to physics-based models. 

\section{A Unified Framework}
\label{sec:model_framework}

Taken all together, our development provides a unified framework for modeling complex
systems from partial observations. Now, with it laid out, we can connect
physics-based and data-driven prediction. We find that the two, seemingly
disparate, paradigms fall at two extremes of the same spectrum: physics-based
models are fully \emph{explicit} and data-driven models are fully
\emph{implicit}.

Recall that physics models reconstruct a coarse-grained data image $u_t =
\overleftarrow{a}(\past^k_t)$ that explicitly fills in missing degrees of
freedom using data assimilation over past observations. The model dynamic
$\simulation$ evolves data images $u_t$ by explicitly computing the
interactions among its degrees of freedom. In this way, the orbits of data
images are generated by:
\begin{align*}
u_{t+\tau} & = \simulation(u_t) \\
  & = \simulation\bigl(\overleftarrow{a}(\past^k_t)\bigr)
  ~.
\end{align*}

This is reminiscent of Hilbert space models
$\overleftarrow{\mathcal{T}}^\tau(\past^k)$, although the above technically
evolves data images. However, due to the explicit nature of data images we can
define \emph{simulated measurements} $\widetilde{x}_t = \widetilde{X}(u_t)$
that produce instrumental readings for the data images. In this way, we predict
instrument readings using physics models via:
\begin{align}
x_{t+\tau} & = \widetilde{M}^\tau_0(\past_t^k) \nonumber \\
  & = [\widetilde{X} \circ \simulation \circ \overleftarrow{a}](\past_t^k)
  ~,
\label{eq:physics-markov}
\end{align}
Again, this is all implemented \emph{explicitly} in
terms of interactions among the observed and inferred unobserved degrees of
freedom.

Equation (\ref{eq:physics-markov}) now clearly parallels the optimal
history-dependent Hilbert space model:
\begin{align}
    x_{t+\tau} &= \markovOP^\tau(\past_t^k) \nonumber \\
    &= [P_W U^\tau X](\past_t^k)
    ~.
\end{align}
Rather than explicitly fill-in the missing degrees of freedom with assimilated
data images, data-driven models use the intrinsic geometry of coordinate
embeddings to implicitly fill-in the missing variables. Whereas physics models
attempt to directly approximate Platonic differential equations-of-motion,
data-driven models attempt to approximate the action of the Platonic Koopman
operator on embedding coordinates via Wiener projections. And, as we
demonstrated, they may converge to the Platonic model in the limit of
infinite-length embeddings.

Markovian closure in their dynamics motivated our introducing Platonic models.
Model parametrizations, though, are used for physics models if the data images
cannot provide adequate closure. Analogously, if a finite-dimensional embedding
does not provide adequate closure, there is still a nonzero orthogonal
component in the Mori-Zwanzig equation resulting from the Wiener projection on
the action of the Koopman operator. In this case, a noise term can be added as
a \emph{stochastic parametrization} to alleviate the lack of closure
\cite{chor15a}.

Between fully-explicit and fully-implicit models lies a spectrum including
history-dependent models that combine implicit and explicit modeling. In
particular, the spectrum encompasses the recent trend in \emph{physics-informed
machine learning} (PIML) \cite{will20a,karn21a,kash21a}. There, known physical
constraints are explicitly incorporated into the model, usually in the form of
conservation laws. The model is then trained from data to implicitly learn the dynamics while maintaining the explicitly enforced constraints.

The unified framework allows clearly evaluating the advantages and
disadvantages of various modeling paradigms. Having explicit access to degrees
of freedom in physics-based models allows for their direct manipulation in the
model. This greatly facilitates, for example, making projections of future
climate outcomes under various anthropogenic forcing scenarios. Such uses of
physics-based models are becoming increasingly common under the heading of
\emph{digital twins} \cite{tao18a,bosch16a}.

That said, on the one hand, difficulties arise with physics models. This is
particularly the case for prediction and especially when confronted with the
poor coverage provided by observed degrees of freedom. Said simply, it is
challenging to construct good data images that approximate the system state
well. Similarly, if there are many important interactions to track, such as in
the climate system, it is impossible to explicitly account for all interactions
and so parametrizations are required. Compounding these problems, generally, it
is not clear how to construct appropriate or effective parametrizations. Due to
all these challenges, implicit approaches are increasingly being added to
physics models, particularly to provide data-driven parametrizations
\cite{schn17a,dura19a}.

On the other hand, though famously difficult to interpret, data-driven models
often excel at straightforward prediction and forecasting tasks. This is no
longer surprising. The unified framework provided a physical explanation for
this success. While data-driven models can converge to the Platonic model in
the limit, however, in practice, they must be learned from finite resources.
Deep learning models, in particular, can be prohibitively computationally
expensive to train. Adding known physical constraints, when applicable, can
help such models converge more quickly. The lesson is that models should not
implicitly learn already-known features; the latter should be incorporated
explicitly.

\section{Conclusion}

Many modeling applications attempt to predict a physical system's future
behavior but can access only a small subset of the system degrees of freedom.
Historically, predictions with partial observations have relied on
physics-based models that explicitly fill-in missing degrees of freedom using
data assimilation and parametrization. In this, the physics models provide
approximate solutions to the governing equations of motion---the system's true
dynamics or Platonic model. Data-driven approaches, in contrast, learn the
dynamics of the observed variables using implicit representations of all the
degrees of freedom through delay-coordinate embeddings.

We demonstrated how the maximal predictive information available to a
data-driven model---information from past observations of the accessible
variables---is given by the predictive distributions. We gave an explicit
construction using Koopman and Perron-Frobenius operators. Most data-driven
models are Hilbert space models, in the form of a target function, that map
past observations forward in time. Maximum Caliber measures were used to
develop a nonequilibrium version of Wiener projections for the Mori-Zwanzig
formalism. Using this, we showed that optimal Hilbert space models correspond
to expectation values of the predictive distributions. Building on the
intuition from generating partitions of symbolic processes, this insight
illuminated how optimal Hilbert space models converge to the true evolution of
the accessible variables, in the limit of infinite-length coordinate
embeddings. We also showed how Wiener projections provide a clear theoretical
connection between data-driven and physics-based models.

At first blush, the empirical success of data-driven models is
counterintuitive. Indeed, by definition, they know nothing of the underlying
physics governing the full system. And yet, they still learn, and from only
partial observations, to predict the true evolution; i.e., they converge to the
Platonic model. Upon reflection, however, we recognize that our understanding
and mathematical formulation of physical laws did not spontaneously manifest.
They formed and evolved over generations precisely through our observations and
interactions with the natural world. The development of science has been
data-driven and successful at that. And so, it is not surprising that
data-driven models ``learn physics'' from observations alone.

What is perhaps discomforting, though, is the implicit and often
uninterpretable manner in which most data-driven methods learn to approximate
the governing physics. We hope that the detailed investigations of implicit
models given here alleviates at least some of this puzzle. It must also be
remembered that initially there was a great deal of discomfort with explicit
numerical physics models. After all, complicated numerical models are very much
``black box'' in ways similar to data-driven models, particularly deep learning
models. The behavior that emerges in complicated physics models often cannot be
deduced directly from the inputs given to that model. If a numerical model
produces unphysical or otherwise pathological behaviors, it is often not
immediately clear how to diagnose and address the concern
\cite{koni92a,carr93a}.


Returning to our motivating question, Is there a \emph{best} way to predict a
given physical system from partial observations? At present, it does not seem
that there is a universally ``best'' approach. When working with finite data
and finite computational resources, all methods have their advantages and
disadvantages. We sought to convey the commonality among seemingly disparate
approaches to predicting complex systems from partial observations. Our goal
was to illuminate the theoretical underpinnings of implicit and explicit
models. Finding commonality in a unified predictive framework should help build
confidence in the models currently employed. And, hopefully, this will pave the
way forward to models with ever more predictive skill and structural
interpretability. At which point, the science of complex systems will have
moved closer to automated theory building \cite{Crut88a}.

\begin{acknowledgments}
We thank Nicolas Brodu, Derek DeSantis, and Jordan Snyder for insightful
discussions, as well as Stefan Klus, Yen Ting Lin, Balu Nadiga, and Dimiter Vassilev for
helpful conversations. The authors also thank the Telluride Science Research
Center for hospitality during visits and the participants of the Information
Engines Workshops there. JPC acknowledges the kind hospitality of the Santa Fe
Institute, Institute for Advanced Study at the University of Amsterdam, and
California Institute of Technology for their hospitality during visits. Part of
this research was performed while AR was visiting the Institute for Pure and
Applied Mathematics, which is supported by the National Science Foundation
grant DMS-1440415. AR acknowledges the support of the U.S. Department of Energy
through the LANL/LDRD Program and the Center for Nonlinear Studies. This
material is also based upon work supported by, or in part by,
APRA E Program: Design Intelligence Fostering Formidable Energy Reduction and Enabling Novel Totally Impactful Advanced Technology Enhancements (DIFFERENTIATE) award number DE-AR0001202,
Templeton World
Charity Foundation grant TWCF0570, Foundational Questions Institute and Fetzer
Franklin Fund grant FQXI-RFP-CPW-2007, U.S. Army Research Laboratory and the
U.S. Army Research Office grants W911NF-21-1-0048 and W911NF-18-1-0028, and
U.S. Department of Energy grant DE-SC0017324.
\end{acknowledgments}

\appendix

\section{Ergodicity and Invariant Measures}
\label{app:ergodicity}

In contrast to conservative Hamiltonian systems---the default assumption for
statistical mechanics---many physical, chemical, and biological systems display
dissipative and nonasymptotic behaviors that demand attention for a full
understanding. We now define these behaviors in detail. This, in turn,
highlights the application breadth of the unified framework.

A system's \emph{phase space} consists of all of its allowed configurations. A
primary goal in dynamical systems theory is to identify the key state-space
structures that guide and constraint a system's complex behaviors
\cite{meiss07a}. We wish to capture them explicitly in our development. This
requires a slightly more general presentation than is usually given for ergodic
theory.

\emph{Invariant sets} are subsets of a system's states that map onto themselves
under a system's dynamic. When perturbations from them return, they are stable
invariant sets---called \emph{attractors}. That set of states which tend
asymptopically to a given attractor is the attractor's \emph{basin of
attraction}. A given dynamical system can be decomposed into its invariant sets
including attractors and their basins and the \emph{basin boundaries}.
Specifying these objects delineates a system's \emph{attractor-basin
portrait}---its comprehensive dynamically-relevant architecture.

Multistable systems are those with multiple attractors. Transient,
nonasymptotic behaviors reflect relaxation to an attractor from states starting
in its basin. The asymptotic stability of attractors meanwhile allows for the
standard long-time analysis of ergodic systems. That is, the standard setup for
the measure-preserving and ergodic dynamical systems of interest to us
describes the evolution on an attractor.

We now formally define these concepts.

Consider the measure space $(\Omega, \Sigma_\Omega, \nu)$ and a dynamic $\Phi$,
where $\Omega$ is the state space, $\Sigma_\Omega$ its Borel algebra, and $\nu$
the Lebesgue measure. A set $A \subset \Sigma_\Omega$ is a
\emph{$\Phi^t$-invariant set} if $\bigl(\Phi^t\bigr)^{-1}(A) = A$ for all $t >
0$, where $\bigl(\Phi^t\bigr)^{-1}(A)$ is the pre-image of $A$ under $\Phi^t$.
In contrast, $A$ is a \emph{forward-invariant} set of $\Phi^t$ if for every
$\omega \in A$, $\Phi^t(\omega) \in A$ for all $t > 0$. Note that all
$\Phi^t$-invariant sets are necessarily also forward-invariant, but not all
forward-invariant sets are $\Phi^t$ invariant.

An attractor of $(\Omega, \Sigma_\Omega, \nu, \Phi)$ is a set $A \subset
\Sigma_\Omega$ with the following properties:
\begin{itemize}
\item $A$ is a forward-invariant set of $\Omega$ under $\Phi^t$,
\item There exists an open set $\mathcal{B} \supset A$, called the basin of
	attraction of $A$ such that for every $\omega \in \mathcal{B}$,
	$\lim\limits_{t \rightarrow \infty} \Phi^t(\omega) \in A$, and
\item There is no proper subset of $A$ with the first two properties.
\end{itemize}

By definition, an attractor is a forward-invariant set. However, due to the
existence of its basin of attraction, an attractor is not a $\Phi^t$-invariant
set. There are points $\omega \in \mathcal{B} \setminus A$ that are in the
pre-image $\bigl(\Phi^t\bigr)^{-1}(A)$ but not in $A$. However, the full basin
of attraction $\mathcal{B}$ for a given attractor $A$ \emph{is}
$\Phi^t$-invariant. (The attractor itself is in its basin $A \subset
\mathcal{B}$.) Every state in $\mathcal{B}$ limits to its attractor $A$. And
so, if there are states in the pre-image of $\mathcal{B}$ that are not in
$\mathcal{B}$ they, by definition, do not limit to $A$. Therefore, any state
not in the pre-image of $\mathcal{B}$ is not in $\mathcal{B}$. In fact, an
alternative definition of the basin $\mathcal{B}$ of attractor $A$ is as the
limit of pre-images of $A$: $\mathcal{B} = \lim\limits_{t \rightarrow \infty}
\bigl(\Phi^t\bigr)^{-1} A$.

Attractors and their basins of attraction decompose a dynamical system into its
dynamically-independent components---the system's attractor-basin portrait. For
a multistable system with multiple attractors, the basins of attraction
partition the state space $\Omega$ into equivalence classes of states based on
the attractor to which they limit since orbits never cross basin boundaries.
Without loss of generality, the development considers dynamical systems with a
single attractor and $\Omega$ its basin of attraction, unless explicitly stated
otherwise. For multistable systems, each attractor and its basin may be
analyzed separately as if it were its own separate system. The full
attractor-basin portrait becomes relevant, though, when one executes
independent experimental trials that select a wide range of initial states.
Moreover, real-world systems are neven fully isolated and this typically
introduces fluctuations that can drive a system between otherwise
noncommunicating basins.

Decomposing a dynamical system into independent components raises the issues
of ergodicity and ergodic measures \cite{laso94a}. A dynamical system $(\Omega,
\Sigma_\Omega, \mu, \Phi)$ is \emph{ergodic} and $\mu$ is an \emph{ergodic
measure}, if every $\Phi^t$-invariant set $B$ is such that $\mu(B) = 1$ or
$\mu(B) = 0$. For an ergodic system, all $\Phi^t$-invariant sets are trivial
subsets of $\Omega$. From the definition of basins of attraction, a dynamical
system with a single basin of attraction or a multi-stable system restricted
to a single basin is ergodic.

Ergodic theory often considers a dynamical system $(\Omega,\Sigma_\Omega, \mu,
\Phi)$ with measure $\mu$ to also be \emph{measure-preserving}:
$\mu\bigl((\Phi^t)^{-1} (B)\bigr) = \mu(B)$ for $B \subset \Sigma_\Omega$. An
equivalent statement is that the measure $\mu$ is \emph{invariant} under the
dynamics $\Phi$.

This is mathematically convenient for casting the behavior of dynamical
processes as stationary stochastic processes. However, it is too restrictive
for our purposes, as it does not capture relaxation to an attractor.  Transient
behavior during relaxation to an attractor $A$ is \emph{dissipative} if it
involves measurable subsets of $\mathcal{B}$ not in $A$, known as
\emph{wandering sets}. In essence, measure is ``carried away'' by wandering
sets, and so the support of an invariant measure cannot include wandering sets.
This can also be seen from the definitions of invariant measures and the
Perron-Frobenius operator above in Eq.~(\ref{eq:perronfrobenius}): a measure is
invariant if and only if it is a fixed point of the Perron-Frobenius operator
\cite[Thm 4.1.1]{laso94a}.

It is often of particular concern whether or not the Lebesgue reference measure
$\nu$ is invariant under the dynamics. Since $\nu$ provides a measure of state
space volume, dynamics that preserve $\nu$ are said to be \emph{volume
preserving}. Wandering sets, by definition, preclude volume preservation.
Hamiltonian systems, on the other hand, are volume preserving due to
Liouville's theorem \cite{Wild98a}. Because we consider only ergodic systems,
there will always be a physical invariant probability measure that may be used,
whether the system preserves volume or not. If the Lebesgue measure is
invariant (and so volume is preserved), the microcanonical distribution gives
the equilibrium invariant probability distribution. If Lebesgue measure is not
invariant, there will be still be a unique asymptotic invariant measure.  (More
on this shortly.) Our formalism works in all cases, but is particularly useful
for generalizing to nonasymptotic behaviors of systems that do not preserve
phase space volume. 

Note that when considering probability measures, the terminology of \emph{measure-preserving} dynamics should not be confused with what we might call \emph{conservation of measure} (or \emph{conservation of probability}). As standard, we assume the dynamics $\Phi$ to be \emph{nonsingular} such that $\mu(\Phi^{-t}(B)) = 0$ for all sets $B$ with $\mu(B) = 0$. This ensures the evolution of probability measures by Perron-Frobenius operators are still probability measures. 

To include transient behavior (relaxation to an attractor), the following does
not assume an invariant measure. However, by restricting to ergodic
dynamics (considering single basins of attraction at a time), this guarantees
the existence of a unique \emph{asymptotic invariant measure}:
\begin{align*}
\mu_{\rho_*} (B) = \int_B \rho_* d\nu
  ~.
\end{align*}
This follows since the $L^1$ Perron-Frobenius operators have a unique invariant
density $\rho^*$ for ergodic dynamics: $P^t \rho_* = \rho_*$ \cite[Thm
4.5]{mack92a}. (These measures play a role roughly analogous to equilibrium
macrostates in thermodynamics.) That is, this measure is preserved by dynamics
on the attractor, to which the system is restricted in the limit. Therefore,
in the asymptotic limit the ergodic theorem applies and time averages equal
state space averages for observables $f$:
\begin{align*}
\lim\limits_{n \rightarrow \infty} \frac{1}{n} \sum_{k=0}^{n-1} f \bigl(\Phi^k (\omega) \bigr) = \frac{1}{\mu_*(\Omega)} \int_\Omega f(\omega) d\mu_*
  ~.
\end{align*}

In the asymptotic limit, the system trajectories settle on the attractor and
the resulting dynamical process is distributed according to the asymptotic
invariant measure $\mu_*(B)$. Thus, by definition, the process is stationary
only in the limit:
\begin{align*}
\Pr(X_t \in B_{\mathcal{X}}) & = \int_{X^{-1}(B_{\mathcal{X}})} d\mu_t^* \\
   & = \int_{X^{-1}(B_{\mathcal{X}})} d\mu_{t+\tau}^* \\
   & = \Pr(X_{t+\tau} \in B_{\mathcal{X}})
  ~.
\end{align*}
Generally, though, $\mu_t \neq \mu_{t+\tau}$.

\section{Optimal Finite-Precision Instruments for Continuous Observables}
\label{app:gen_part}

The following temporarily leaves behind the fully-continuous dynamical
processes setting. Instead, it considers discrete-time, discrete-valued
\emph{symbolic processes} \cite{Lind95a} and how they relate to discrete
measurements of continuous dynamical systems. In this, the mapping $X$
corresponds to a coarse-grain partition $\partition$ of the state space
$\Omega$. As with dynamical processes, $X$ is many-to-one and noninvertible,
yielding fully-discrete stochastic processes of observations. Rather than
interpreting $X$ as accessing only a subset of accessible degrees of freedom in
$\omega$, for symbolic processes $X$ is interpreted as a collection of
measurement instruments, each with access to all relevant degrees of freedom,
but only report the result of finite-precision observations \cite{Casd91a}. To
avoid confusion, we denote the observation function for symbolic processes as
$\Xp$.

\subsection{Symbolic Processes from Generating Partitions}

A symbolic measurement function $\Xp: \Omega \rightarrow \mathcal{A}$ generates
a finite partition $\partition$ of state space $\Omega$, with every $\omega \in
\Omega$ mapping to a partition element $\partition_i$ such that $\partition_i
\cap \partition_j = \emptyset$ for all $\partition_i, \partition_j \in
\partition$ and $\bigcup_i^K \partition_i = \Omega$. Each partition element
$\partition_i$ carries a label, or \emph{symbol} $a_i \in \mathcal{A}$. Without
loss of generality, we will take label$(\partition_i) = i$, with $\mathcal{A} =
\{0, 1, \ldots, K-1\}$ for $K$ partition elements in $\partition$. Using this,
we can explicitly write the piecewise constant symbolic measurement function
$\Xp$ in terms of the partition elements as:
\begin{align}
\Xp (\omega) = \sum\limits_{i=0}^{K-1} i \mathds{1}_{\partition_i} (\omega)
  ~,
\label{eq:symb_meas}
\end{align}
where:
\begin{align*}
\mathds{1}_{\partition_i} (\omega) =
   \begin{cases}
   1 & \omega \in \partition_i \\
   0 & \omega \notin \partition_i
   \end{cases}
\end{align*}
is the indicator function for partition element $\partition_i$.


Paralleling our development of dynamical processes, we now consider symbol
sequences generated by measuring orbits of the underlying system $(\Omega,
\Phi)$. For an initial value $\omega_0$ there is an initial symbol $a_0 =
\Xp(\omega_0)$---an element of partition $\partition$. Similarly, $\Xp \circ
\Phi$ induces a partition over $\Omega$, denoted $\Phi^{-1} \partition$, such
that each element $(\Phi^{-1} \partition)_i$ is the set of all $\omega$ for
which $\Xp \bigl(\Phi(\omega)\bigr) = \partition_i$. That is, $\Phi^{-1}
\partition$ is a partition over $\Omega$ at the initial time $t_0$ where every
$\omega_0$ in the same element of $\Phi^{-1} \partition$ emits the same symbol
$a_1 = \Xp(\omega_1) = \Xp\bigl(\Phi(\omega_0)\bigr)$ at the next time $t_1$.
Each time step $t_n$ generates a new partition $\Phi^n \partition$ whose
elements are all the points $\omega_0 \in \Omega$ such that $\Xp
\bigl(\Phi^n(\omega_0)\bigr) \in \partition_i$.

Importantly, an iterated partition \emph{refines} the previous partition. For
two partitions $\partition$ and $\mathbb{Q}$, the \emph{refinement} $\partition
\vee \mathbb{Q} = \{\partition_i \cap \mathbb{Q}_j$, for all $\partition_i \in
\partition$ and $\mathbb{Q}_j \in \mathbb{Q}\}$ is also a partition. The first
refinement of $\partition$ under $\Phi$ is $\partition \vee \Phi^{-1}
\partition$. Its elements are all the points $\omega_0 \in \Omega$ that emit
the same symbol $\Xp(\omega_0)$ for time $t_0$ and that emit the same symbol
$\Xp\bigl(\Phi(\omega_0)\bigr)$ at the next time $t_1$. Therefore, the
refinement $\partition \vee \Phi^{-1} \partition$ maps from $\Omega$ to
two-symbol sequences $a_0 a_1$ in $\mathcal{A} \times \mathcal{A}$. In the
limit, the full dynamical refinement $\partition \vee \Phi^{-1}\partition \vee
\Phi^{-2}\partition \vee \cdots$ maps points in $\Omega$ to infinite-length
symbol sequences in $\mathcal{A} \times \mathcal{A} \times \mathcal{A} \times
\cdots$.

A partition $\partition$ is \emph{generating} if there is a one-to-one
correspondence, almost everywhere, between an initial condition $\omega_0 \in
\Omega$ and the infinite sequence of symbols $\{\Xp(\omega_0),
\Xp\bigl(\Phi(\omega_0)\bigr), \Xp\bigl(\Phi^2(\omega_0)\bigr), \ldots\}$
generated by $\omega_0$. Thus, while the initial measurement symbol $a_0 =
\Xp(\omega_0)$ is far from sufficient to fully determine $\omega_0$, the full
infinite sequence of subsequent symbols \emph{does} (almost-everywhere) fully
determine $\omega_0$ if $\partition$ is a generating partition. This occurs
since the size of the dynamical refinement partition elements goes to zero in
the infinite-time limit. And, in turn, this requires the system to be chaotic; exponential spreading of orbits in forward time corresponds to exponential convergence in reverse time.

Due to all this, generating partitions provide a rigorous notion of a ``good''
measurement device for which information lost by a coarse single-time
measurement is recovered through an infinite-time limit of measurement
observations. The one-to-one correspondence property of generating partitions
emerges above when discussing the potential convergence of data-driven models of
partially-observed systems.
The set $\overleftarrow{B}_t^k$ is
the element of the dynamical refinement of the generating partition
corresponding to the observed symbol sequence $\past_t^k$. In the limit of infinitely-many
observations the size $\nu(\overleftarrow{B}_t^k)$ of the refined partition elements vanishes and
almost-every infinite-length symbol sequence corresponds to a unique system
state $\omega \in [0,1]$.

An important bridge between symbolic and dynamical processes arises from the fact that the evolution of partitions
$\Phi^{-n}\partition$ (not the dynamical refinements) is governed by
discrete-time Koopman operators. The partition $\Phi^{-n}\partition$ is
generated by the time-shifted symbolic measurement function $\Xp \circ \Phi^n =
U^n \Xp$. Therefore, again paralleling dynamical processes, the symbol
sequences are given by $\{\Xp(\omega_0), [U \Xp](\omega_0), [U^2
\Xp](\omega_0), \ldots\}$.



\begin{figure}
\begin{center}
\includegraphics[width=0.48 \textwidth]{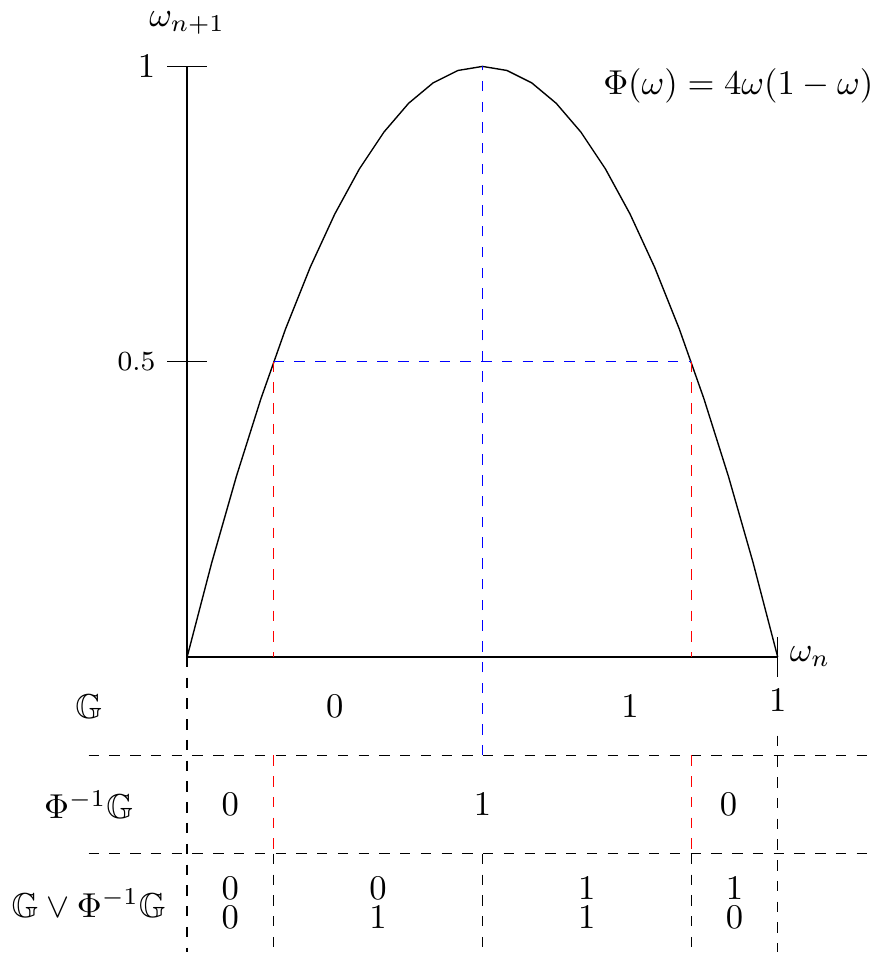}
\end{center}
\caption{Logistic map of the unit interval at $r=4$: Shown with generating
	partition $\mathbb{G}$, the single-time evolved partition
	$\Phi^{-1}\mathbb{G}$, and the first dynamical refinement partition
	$\mathbb{G} \vee \Phi^{-1}\mathbb{G}$.
	}
\label{fig:logistic}
\end{figure}

\subsection{Generating Partition of the Logistic Map}

A common arena for investigating symbolic processes of chaotic dynamical
systems considers continuous maps on the unit interval $\Omega = [0,1]$
\cite{Miln77,Coll80a}. Here, we examine the logistic map:
\begin{align*}
\omega_{n+1} & = \Phi(\omega_n) \\
  & = r\omega_n(1-\omega_n)
  ~,
\end{align*}
We set $r = 4$.

The binary partition $\mathbb{G}$, shown in Fig.~\ref{fig:logistic}, with
$\mathbb{G}_0 = [0, \frac{1}{2}]$ and $\mathbb{G}_1 = [\frac{1}{2}, 1]$, is a
generating partition of the logistic map \cite{Coll80a}. The corresponding
symbolic measurement function is the step function:
\begin{align*}
X_{\mathbb{G}}(\omega) =
	\begin{cases}
	0 & 0 \leq \omega \leq 0.5 \\
	1 & 0.5 \leq \omega \leq 1
	\end{cases}
    ~.
\end{align*}
Note this function is in the general form of Eq.~(\ref{eq:symb_meas}).

The single-time evolved partition $\Phi^{-1} \mathbb{G}$, also shown in
Fig.~\ref{fig:logistic}, is given by the single-time shift symbolic measurement
function:
\begin{align}
X_{\mathbb{G}}\bigl(\Phi(\omega)\bigr) & = [U X_{\mathbb{G}}](\omega) \\
  & = \begin{cases}
  1 & \frac{1-\sqrt{\frac{1}{2}}}{2} \leq \omega
  \leq \frac{1+\sqrt{\frac{1}{2}}}{2} \\
  0 & \mathrm{otherwise}
  \end{cases}
    ~.
\end{align}
The new boundary points $\left( 1-\sqrt{\frac{1}{2}}\right)/2$ and $\left(
1+\sqrt{\frac{1}{2}}\right) / 2$ of $\Phi^{-1}\mathbb{G}$ are the pre-images
$\{ \Phi^{-1} (\omega) \}$ of the original boundary point $\omega = 1 / 2$ of
$\mathbb{G}$.

Finally, the first dynamical refinement $\mathbb{G} \vee \Phi^{-1}\mathbb{G}$,
mapping $\Omega$ to two-symbol sequences, is also shown in
Fig.~\ref{fig:logistic}. From Fig.~\ref{fig:logistic} we see that the dynamical
refinement adds the boundary points of $\Phi^{-1}\mathbb{G}$ to the original
boundary point of $\mathbb{G}$.

\bigskip

Beyond rigorously formulating good measurement devices---generating
partitions---symbolic processes were historically important for introducing
concepts and methods from discrete information and computation theories into
dynamical systems and ergodic theory, as noted above. In particular, the
Shannon entropy rate of a symbolic process has a (possibly nonunique) supremum
over all possible partitions for a given iterated map. This is the
\emph{Kolmogorov-Sinai entropy}. That is, the supremum is achieved for
generating partitions~\cite{kolm59a,sina59a}. Moreover, the Kolmogorov-Sinai
entropy is bounded by the positive Lyapunov exponents of the underlying system
\cite{pesi77a}. This provides a rigorous link between the geometric
instabilities of deterministic chaos and observed randomness. And, this
explains, in part, why the weather is hard to predict \cite{Lore63a,Lore64a}.

\section{Data-Driven Koopman Approximation}
\label{app:koopman_appox}


Given that $U^\tau$ provides the ground-truth for $\iddmodel$, why not use a
data-driven approximation of $U^\tau$ for $\iddmodel$? Finite-dimensional
approximations of $U^t$ are useful for global spectral analysis of nonlinear
systems, but they are typically not optimal for predictive modeling, as we
now show.

\begin{figure*}
\begin{center}
\includegraphics[width=1.0 \textwidth]{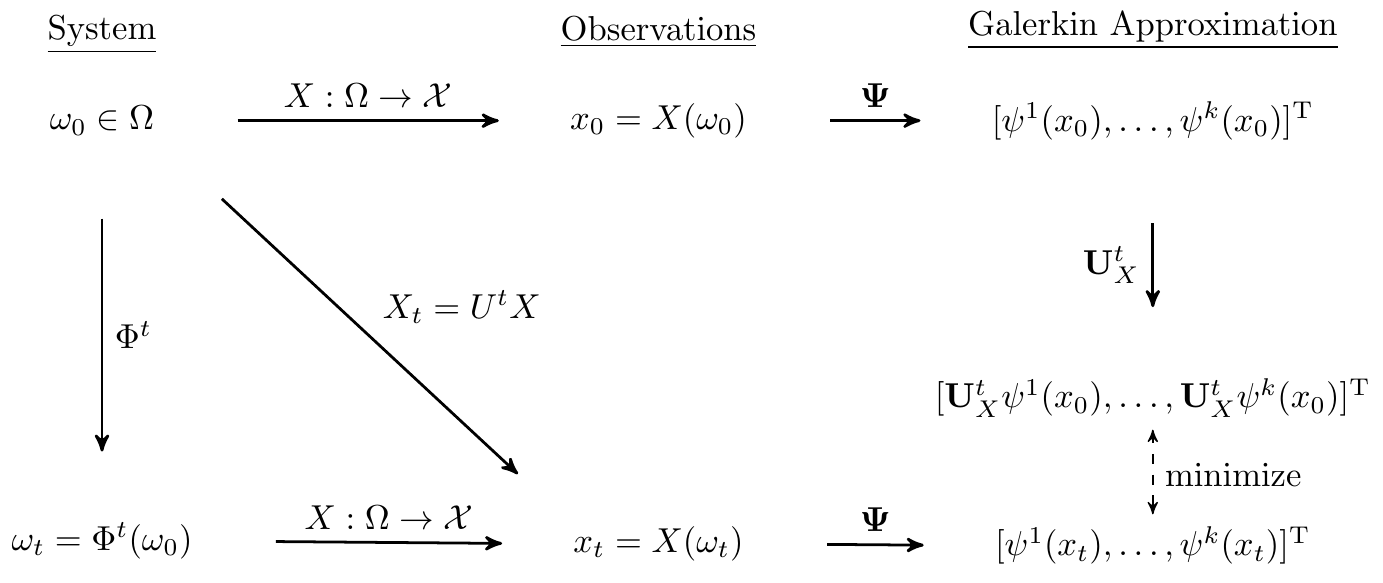}
\end{center}
\caption{EDMD algorithm's commuting diagram for finite-dimensional Galerkin
	approximation $\edmd$ of $U^t$ onto $H_\dict \subseteq H_X \subset H$.
	}
\label{fig:EDMD}
\end{figure*}

Data-driven finite-dimensional---i.e., matrix---approximations of $U^\tau$ are
most generally understood through the \emph{Extended Dynamic Mode
Decomposition} (EDMD) algorithm \cite{will15a,klus16a} shown in
Fig.~\ref{fig:EDMD}. Consider a dictionary $\dict = [\psi^1(x), \ldots,
\psi^k(x)]^{\text{T}}$ of basis functions in $V$. For simplicity, assume this
is an orthonormal set so that $\dict$ defines the closed Hilbert subspace
$H_{\dict} \subseteq H_X \subset H$ spanned by $\dict \circ X$. Given a set of
training data $\timeseries$, EDMD finds a (least squares) best-fit matrix
$\edmd$ such that:
\begin{align}
    \dict(x_{t+\tau}) = \edmd \dict(x_{t})
  ~.
\end{align}
This is typically an overdetermined optimization, and so it is common to pick a
solution by applying the pseudoinverse $\dict^+$, giving:
\begin{align}
[\edmd]^{\text{T}} = \dict(x_{t+\tau}) \dict^+(x_t)
    ~.
\end{align}
In the infinite data limit, $\edmd$ converges to a Galerkin projection of
$U^\tau$ onto $H_\dict$, so that:
\begin{align}
    \langle \psi_j, U^\tau \psi_i \rangle = \langle \psi_j, \edmd \psi_i \rangle
	~,
\end{align}
for all $i,j = 1,\ldots, k$.

In the fully-observed case, where $X$ is the identity ($X(\omega) = x =
\omega$), the Galerkin projection $\mathrm{\mathbf{U}}^\tau$ converges to the
true Koopman operator $U^\tau$ in the limit of an infinitely-large dictionary
$\dict$, where $H_\dict \rightarrow H$ \cite{kord18a}. However, in the
partially-observed case, the dictionary is restricted to functions of partial
observations $x$ only. Thus, in the limit of an infinitely-large dictionary
$\dict \rightarrow V$, we only have that $H_\dict \rightarrow H_X$. Therefore,
$\edmd$ cannot converge to the full $U^\tau$. We include the subscript $X$ in
$\edmd$ to signify this fundamental restriction.

Several difficulties arise in using $\edmd$ as a predictive model. First and
foremost, the identity function $f_X(x) = x$ must be included in $\dict$.
($f_X$ is sometimes called the \emph{full-state observable} in the Koopman
literature, but we do not as it is confusing in the setting of
partially-observed systems.) A prediction is then given as:
\begin{align*}
x_{t+\tau} & = \iddmodel_{\text{EDMD}}(x_t) \\
  & = \edmd [f_X \circ X] (\omega_t)
    ~.
\end{align*}
That is, the forecast is determined by the action of $\edmd$ on the identity
observable $f_X \circ X$.

To be clear, $\edmd$ is an operator on the Hilbert subspace $H_\dict \subseteq
H_X \subset H$ of observables of the full underlying system $\Omega$. However,
due to the partial-observation constraint it must always act on observables
composed with $X$. (And so, it can be thought of as acting on functions of
$x$.) However, the identity observable $f_X$ need not be included in
constructing the dictionary $\dict$. The constraint of requiring $f_X \in
\dict$ can be avoided through the use of autoencoder neural networks to
construct $\edmd$ \cite{li27a,lusc18a,otto19a}. The decoder of the network
learns a nonlinear map from $H_\dict \to \mathcal{X}$ that recovers $f_X$ as a
nonlinear combination of the elements of $\dict$.

In practice, the distinction between discrete-time and continuous-time systems
can be important. For continuous time, the gEDMD algorithm \cite{klus20a}
should be employed to approximate the Koopman generator. This is done by using
finite differences or automatic differentiation of the observation time series.

A more serious difficulty in using EDMD for prediction comes from its its lack
of closure---leakage out of the subspace $H_\dict$. If $H_\dict$ is not a
finite Koopman-invariant subspace \cite{brun16a}, then after several iterations
$U^\tau[f_X \circ X]$ eventually no longer lies within $H_\dict$. Due to this,
$\edmd$'s action differs from the true evolution given by $U^\tau$'s action.
Note that if $H_X$ is not a Koopman invariant subspace, then \emph{all}
instantaneous models $\iddmodel$ accrue a similar prediction error. This is the
intrinsic error $\Xi^\tau_X$ discussed above, which is incurred for having only
partial observations $X$ of $\Omega$.

In the infinite dictionary limit, $\dict \rightarrow V$ and so $\edmd [f_X
\circ X] = g \circ X$ is always in $H_\dict = H_X$, for some $g \in V$. Thus,
EDMD converges in the limit to optimal target function---regression
function---$Z^\tau$ in Eq.~(\ref{eq:target_estimator}). Given that the Koopman
operator provides the ground-truth for data-driven models, it is not surprising
that the EDMD approximation method for $U^\tau$ recovers the optimal instantaneous target
function.

The difficulty is that EDMD never reaches the $\dict \rightarrow V$ limit.
Therefore, generally the leakage of $\edmd [f_X \circ X]$ out of $H_\dict$ may
still lie within $H_X$. Unlike the $\dict \rightarrow V$ limit, with leakage
out of $H_X$, this leakage is avoidable, given a better choice of or larger
dictionary $\dict$. Moreover, the prediction error from the leakage compounds
over time. The choice of $\dict$ thus substantially impacts EDMD's predictive
skill. In general, a finite invariant subspace cannot be determined a priori.
And, for that matter, may not exist for a given physical system $\Omega$ with a
given $X$---the set of measurements that can be made on $\Omega$. Recent deep
learning approaches \cite{li27a,lusc18a,otto19a,guli21a} attempt to learn an
optimal $\dict$ from data. Similarly, kernel methods
\cite{will15b,klus20b,das20a} are used to create a very large, implicitly-defined, dictionary.

Note that employing the trivial dictionary $\dict = \{f_X\}$,
which includes only the identity, yields the exact Dynamic Mode Decomposition
(DMD) algorithm \cite{tu14a}. For prediction DMD finds the optimal matrix (i.e. linear)
solution for $\iddmodel$ which minimizes the instantaneous target function error in Eq.~(\ref{eq:iddmodel}).

For complex, nonlinear systems, using a linear model for prediction may seem
like a bad idea. Interestingly, though, ``linear plus noise'' models, such as
Linear Inverse Modeling (LIM) \cite{tu14a}, can be reasonably effective and are
frequently used in climate science \cite{alex08a}. We are not aware of attempts
to generalize this to an Extended Linear Inverse Model that implements EDMD
plus noise. The efficacy of LIM models suggests the tolerance induced by noise
may help alleviate the effects subspace leakage.

\paragraph*{Equations of Motion From Data}
A popular approach for data-driven modeling learns an explicit closed-form
equation model for $\iddmodel$. This is referred to as \emph{equation
discovery}. The most common approach performs a dictionary regression;
sometimes also called \emph{symbolic regression} \cite{Crut87a,brun16b}. Like
EDMD, a dictionary $\dict = [\psi^i, \ldots, \psi^k]$ of functions is chosen
and a (typically sparse) regression is performed to find the best-fit
coefficients $a_i$ that minimize $\|\dot{x}_t - \sum_i a_i \psi^i(x_t)\|^2$. In
fact, the dictionary regression approach to equation discovery is a special
case of gEDMD \cite{klus20a}.

Whatever form of equation discovery is used, the ultimate goal is to
approximate $\dot{x} = \Phi_X (x)$ with a closed-form expression for $\Phi_X$.
For partially-observed dynamics, though, it is not guaranteed that $\Phi_X$
will be well-represented by closed-form equations of motion, even if $\Phi$ is
\cite{Crut87a}. The insight that dictionary regression is a special case of the
gEDMD algorithm for approximating the Koopman generator illustrates that
forcing $\iddmodel$ to be closed-form is an unnecessary restriction. There are
certainly many advantages to having closed-form models, including
interpretability and extracting adjustable physical parameters. In contrast,
for prediction our unified framework demonstrates that it is often advantageous
to use \emph{implicit} models for $\iddmodel$.

\bibliography{ergodic}

\end{document}